\documentclass[reprint,aps,prd,notitlepage,superscriptaddress,nofootinbib]{revtex4-2}

\usepackage[utf8]{inputenc}
\usepackage[english]{babel}
\usepackage{amsmath}
\usepackage{amsfonts}
\usepackage{amssymb}
\usepackage{listings}
\usepackage{lipsum}
\usepackage{multirow}
\usepackage{datetime}
\usepackage{graphicx}
\usepackage{mathtools}
\usepackage{mathrsfs}
\usepackage{dcolumn}
\usepackage{multirow}
\usepackage{subfig}
\usepackage{soul}
\usepackage{color} 
\usepackage{hyperref}
\hypersetup{
    colorlinks=true, 
    pdfborder = {0 0 0.5 [3 3]},
    anchorcolor=black,
    citecolor=blue,
    linktoc=all,    
    linktocpage=true,
    linkcolor=red,
	urlcolor=blue
}

\begin{document}

\title{Nonlinear effect of absorption on the ringdown of a spinning black hole}

\author{Taillte May}
\email{tmay@perimeterinstitute.ca}
\affiliation{Perimeter Institute for Theoretical Physics, Waterloo, Ontario N2L 2Y5, Canada}
\affiliation{Department of Physics \& Astronomy, University of Waterloo, Waterloo, Ontario N2L 3G1 Canada}

\author{Sizheng Ma}
\email{sma2@perimeterinstitute.ca}
\affiliation{Perimeter Institute for Theoretical Physics, Waterloo, Ontario N2L 2Y5, Canada}

\author{Justin L. Ripley}
\email{ripley@illinois.edu}
\affiliation{Illinois Center for Advanced Studies of the Universe, Department of Physics, University of Illinois Urbana-Champaign, Urbana, Illinois 61801, USA}

\author{William E. East}
\email{weast@perimeterinstitute.ca}
\affiliation{Perimeter Institute for Theoretical Physics, Waterloo, Ontario N2L 2Y5, Canada}

\date{\today}

\begin{abstract} 
The ringdown gravitational wave signal arising e.g., in the final stage of a black
hole binary merger, contains important information about the properties of
the remnant, and can potentially be used to perform clean tests of general
relativity. However, interpreting the ringdown signal, in particular when it is the loudest,
requires understanding the role of nonlinearities and their potential impact on modeling this phase
using quasinormal modes. Here, we focus on a particular nonlinear effect
arising from the change in the black hole's mass and spin due to the
partial absorption of a quasinormal perturbation. We isolate and
systematically study this third-order, secular effect by evolving the
equations governing linear metric perturbations on the background of a
spinning black hole, but allowing the properties of the background to
evolve in a prescribed way. We find that this leads to the excitation of
quasinormal modes with higher polar angular number, retrograde modes
(counterrotating with respect to the black hole), and overtones, as well
as giving rise to a component of the signal at early times that cannot be
fully described using quasinormal modes. Quantifying these effects, we
find that they may be relevant in analyzing the ringdown in black hole
mergers.

\end{abstract}

\maketitle

\section{Introduction}

In the ultimate phase of a black hole merger, the remnant black hole
relaxes to its final state, emitting a ringdown gravitational
wave signal. As the system transitions from the nonlinear and highly dynamical merger phase,
it eventually becomes well approximated by linear gravitational perturbations of a Kerr black hole,
characterized by a superposition of damped sinusoids called
quasinormal modes \cite{Teukolsky:1973ha,Kokkotas:1999bd,Berti:2009kk}%
\footnote{In this work, we will not consider tails, which can dominate at very late times.}. 
There are several motivations for analyzing the ringdown part of the gravitational wave signal. The
frequencies and decay rates of these quasinormal modes are completely
determined by the properties of the remnant black hole \cite{Carter:1971zc},
and are independent of the form of the initial perturbation. Thus, measuring this
behavior in a gravitational wave signal would be a strong test of consistency
with general relativity \cite{Berti:2018vdi,Cardoso:2019rvt,Dreyer:2003bv}, as
well as an independent measurement of the mass and spin of the black hole
remnant. They can also be used to probe the environment for matter \cite{Bamber:2021knr}.  Measuring
these modes accurately from gravitational wave observations is an active area of
research
\cite{Cotesta:2022pci,Isi:2019aib,LIGOScientific:2020tif,CalderonBustillo:2020rmh,Isi:2021iql,Carullo:2019flw,Finch:2022ynt}. 

An important aspect both of modeling and physically interpreting the ringdown gravitational wave signal
is determining the form and magnitude of nonlinear effects. Understanding in what
regime such effects are significant determines how accurate a purely linear analysis in 
terms of quasinormal modes will be at different stages of the ringdown; it also has the potential to unlock
new aspects of the ringdown signal that might be observed.
A number of recent works have addressed different aspects of this problem using tools from numerical relativity
and beyond linear-order black hole perturbation theory.

At second-order in perturbation amplitude, linear quasinormal modes combine to
generate quadratic quasinormal modes. Analyses of numerical relativity
simulations of binary black hole mergers have found evidence for the presence
of these modes~\cite{Cheung:2022rbm,Mitman:2022qdl,Ma:2022wpv,Cheung:2023vki}, while second-order perturbation calculations have sought to determine
the relation between the amplitude of the quadratic and linear modes
\cite{Nakano:2007cj,Redondo-Yuste:2023seq,Zhu:2024rej,Ma:2024qcv,Bucciotti:2024zyp,Bourg:2024jme}.
Including these modes in a fit to the ringdown can improve the extraction of
the linear modes \cite{Cheung:2023vki,Giesler:2024}.

Attendant in the decay of a quasinormal mode is a flux of energy and angular momentum
through the black hole horizon.  At third order, the resulting change in the
black hole's spin and mass will affect the original perturbation, including by
exciting additional modes through the background dynamics---an effect sometimes
referred to as absorption induced mode excitation (AIME)
\cite{Sberna:2021eui,Redondo-Yuste:2023ipg,Zhu:2024dyl}.  While this effect is
subdominant to quadratic effects in a na\"ive order counting, it is cumulative
with time and has been shown to contribute significantly in some scenarios. In
Ref.~\cite{Sberna:2021eui}, fully nonlinear calculations of scalar quasinormal
modes of spherical black holes in asymptotically anti-de Sitter spacetimes
showed this to be the dominant nonlinear effect in that context, and motivated analytic
estimates of AIME for overtones in asymptotically flat, spherically symmetric spacetimes,
suggesting this effect would play a role in black hole mergers.  In
Ref.~\cite{Redondo-Yuste:2023ipg}, the effect of a changing black hole mass in
spherical symmetry was modelled with null dust, and a model to capture the
resulting changing frequency and amplitude of a ringdown signal was proposed.
Recently, Ref.~\cite{Zhu:2024dyl} studied how a spinning black hole was
perturbed by an incoming gravitational wave packet as a function of its
amplitude using nonlinear simulations. They quantified the effect of the change
in the black hole's mass and spin resulting from the incoming gravitational
wave packet on the amplitude and frequency of the ensuing ringdown signal.

In this work, we also consider the nonlinear effect of a changing black hole
mass and spin, but focus directly on the backreaction that the least-damped
quasinormal mode of a spinning black hole will have on itself through this
channel. We calculate this effect in isolation by evolving the linearized
Einstein equations in terms of the Weyl tensor---i.e., the Teukolsky
equation---but on the background of a black hole whose mass and spin changes
according to the expected flux of energy and angular momentum through the
horizon. Aided by the choice of hyperboloidal slicing, we use initial data
corresponding exactly to a single quasinormal mode and quantify the resulting
change in the gravitational wave signal due to its partial absorption by the
black hole. While this analysis neglects other nonlinear effects, it allows us
to identify and quantify the impact of one particular effect in isolation and
determine in what regime it will impact the consistency of a purely linear
treatment. We find that, in addition to changing the amplitude and phase of
the original least-damped quasinormal mode, this excites overtone, higher
angular number, and retrograde modes.  The amplitudes found for these excited
modes are comparable, in some cases, to those found in the ringdown of black
hole mergers, as well as to quadratic quasinormal modes. 

The remainder of this paper is organized as follows. In
Sec.~\ref{sec:quasi-normal_mode_absorption}, we calculate, to leading order in
perturbation theory, the change in mass and spin of a perturbed black hole due
to the partial absorption of a gravitational perturbation, including for the
specific case of a pure quasinormal mode. In Sec.~\ref{sec:methods}, we
describe our methods for constructing quasinormal mode initial data,
numerically evolving linear gravitational perturbations on a changing black
hole background, and identifying and quantifying quasinormal modes in the
resulting gravitational waves. In Sec.~\ref{sec:Results}, we present our main
results, including identifying which new modes are excited and their relative
amplitudes. We also find evidence for a component in the ringdown signal at
earlier times that cannot be well described by a quasinormal mode, which we
examine and compare to a changing frequency
model in Sec. ~\ref{sec:nonmode_content}. In Sec. ~\ref{sec:discussion}, we discuss our results, including
their relevance to black hole mergers, and conclude.
In the appendixes, we provide details on the analytic calculation
of the change in black hole mass and spin due to a solution of the Teukolsky equation (Appendix~\ref{App:analytic_calc_dA}),
discuss some the subtleties in connect the gravitational wave signal with a gravitational perturbation near
a black hole (Appendixes~\ref{App:motivation_perturbing_amp} and~\ref{App:propagation_times}), 
compare different choices for quasinormal mode initial data (Appendix~\ref{sec:Physical_case}),
and provide details on numerical errors and convergence (Appendix~\ref{sec:num_convergence}).

\section{Analytic quasinormal mode absorption}
\label{sec:quasi-normal_mode_absorption}

Our notation is as follows. 
We set $G=c=1$. 
We use superscripts to refer to the order in black hole perturbation theory (e.g. $f^{(2)}$ denotes a second-order perturbation of $f$).
As is common in applications of the Newman-Penrose (NP) formalism, we use the $+---$ metric signature.
The black hole mass is $M$ and its dimensionful spin is $a$ (such that $a/M$ is dimensionless).
We use $\mathfrak{R}/\mathfrak{I}$ to denote the real/imaginary parts of complex numbers e.g., $\mathfrak{R}\left(1+2i\right)=1$.

\subsection{Flux for a frequency-space perturbation}

Here we briefly review how a black hole's mass and spin change due to a linear
gravitational perturbation \cite{Hawking:1972hy,Teukolsky:1974yv}, as calculated using the NP
formalism \cite{Newman:1961qr}. We provide more details of the calculation,
as well as a review of the literature, in Appendix~\ref{App:analytic_calc_dA}. 

We describe linearized gravitational waves via the linearized NP scalars $\Psi_4^{(1)}$ and $\Psi_0^{(1)}$. 
Linear gravitational perturbations can be absorbed by the black hole, which changes its mass and spin.
To find a quantitative relation between the linearized gravitational field and the black hole mass, we follow an approach described in Ref.~\cite{Hawking:1972hy}. First, from the NP equations, we relate $\Psi_0^{(1)}$ to the \emph{second} order perturbation of the NP scalar $\rho^{(2)}$. 
The surface integral of this quantity on the black hole horizon then gives the change in the black hole area $A$.
In ingoing Kerr-Schild coordinates, the relation is
\begin{align}
   \label{eq:eq_for_A2_hh_tetrad}
      \frac{dA^{(2)}}{dv}
      =
      -
      2\int d\varphi d\theta \sqrt{s} \rho^{(2)}\left(v,\vartheta,\varphi\right)
      ,
\end{align}
where $s$ is the induced metric on the black hole horizon.
The integral over the azimuthal angle $\varphi$ and polar angle $\vartheta$ is evaluated at the horizon radius $r_+=M + \sqrt{M^2-a^2}$.
Following Ref.~\cite{Teukolsky:1974yv}, we perform a Fourier decomposition of all perturbed quantities according
to their frequency $\omega$ and azimuthal number $m$, e.g., 
\begin{align}
    \Psi_0^{(1)}\left(v,r,\vartheta,\varphi\right)
    =
    \int \frac{d\omega}{2\pi} e^{-i\omega v}
    \sum_m e^{im\varphi} \tilde{\Psi}_0^{(1)}\left(\omega,r,\vartheta,m\right)
    .
\end{align}
For a given (fixed) $\omega$ and $m$, we can calculate $dA^{(2)}/dv$ (see Appendix~\ref{sec:change_mass_spin_qnm}).
In addition, the first law of black hole thermodynamics relates the change in black hole area to the change in the mass and spin of the black hole (see Appendix \ref{sec:change_mass_spin}). 
We then have~\cite{Teukolsky:1974yv} 
\begin{align}
   \label{eq:mass_change_from_area_change}
    \frac{d^2M}{dvd\Omega} = 
    \frac{\sqrt{M^2-a^2} | \omega | ^2}{8 \pi  \left(2 M r_+| \omega | ^2 -a m \mathfrak{R}\omega 
   \right)}
    \frac{d^2A}{dvd\Omega},
\end{align}
and
\begin{align}
   \label{eq:a_change_from_area_change}
      \frac{d^2a}{dvd\Omega}
      &=\frac{\sqrt{M^2-a^2}  \left(a | \omega | ^2+m \mathfrak{R}\omega\right)}{8 \pi M \left(2 M r_+ | \omega | ^2 -a m \mathfrak{R}\omega 
   \right)}
      \frac{d^2A}{dvd\Omega}
      .
\end{align}
In short, for a given quasinormal mode solution for $\Psi_0^{(1)}$, we can find the corresponding differential change in the black hole mass and spin. 

Next, we summarize the change in mass and spin of a black hole due to the presence of a single quasinormal mode. 
We consider a quasinormal mode solution to $\Psi_0$: 
\begin{align}
   \Psi_0^{(1)}\left(v,r,\vartheta,\varphi\right)
   =
   \mathcal{A}
   \times
   e^{im\varphi - i\omega v}
   R\left(r\right) S \left(\vartheta\right)
   ,
\end{align}
where $R(r)$ is normalized to unity on the horizon, $R(r_+)=1$. 
Using Eq.~\eqref{eq:eq_for_A2_hh_tetrad} and the relevant NP equations (see Appendix \ref{sec:change_mass_spin_qnm} for details), we find that 
\begin{align}
\label{eq:area-change-psi0-mode-sol}
\frac{dA^{(2)}}{dv}
    = 4\pi
      \left|\mathcal{A}\right|^2
      \frac{ M \mathrm{exp}\left[
            2\left(\mathfrak{I}\omega\right)v
         \right] r_+  }
         { \left(\epsilon^{(0)}_{(HH)}-\mathfrak{I}\omega \right) \left(\mathfrak{R}\omega^2 + ( 
     2 \epsilon^{(0)}_{(HH)}-\mathfrak{I}\omega)^2\right)} 
     ,
\end{align}
where $\epsilon^{(0)}_{(HH)}$ is the NP scalar $\epsilon$ in the Hawking-Hartle tetrad on the horizon [see Eq.~\eqref{eq:epsilon_def}]. 
This result matches others in literature [Eqs. (5.14) and (5.22) in Ref.~\cite{Poisson:2004cw}, as well as \cite{Teukolsky:1974yv} and \cite{Sberna:2021eui}] for a normal mode (i.e. $\mathfrak{I}\omega = 0$) up to coordinate and tetrad changes.\footnote{
Note that our result disagrees with Ref.~\cite{Sberna:2021eui} slightly for the complex mode case. 
The factor of $\left(\epsilon^{(0)}_{(HH)}-\mathfrak{I}\omega \right)$ in the denominator of Eq.~\eqref{eq:area-change-psi0-mode-sol} is replaced by $\left(\epsilon^{(0)}_{(HH)}\right)$ in Ref.~\cite{Sberna:2021eui}. 
}

\subsection{Mass and spin change due the absorption of the least-damped mode} 
\label{sec:mass_spin_change_due_to_qnm}

In order to connect a quasinormal mode, as measured through a gravitational wave signal, to an induced change in the black hole's mass
and spin, we need to relate the amplitude of the quasinormal mode at null infinity to the amplitude of the gravitational perturbation 
on the horizon. This will then allow us to apply Eqs.~\eqref{eq:mass_change_from_area_change}, \eqref{eq:a_change_from_area_change}, and~\eqref{eq:area-change-psi0-mode-sol}.
In Appendix~\ref{App:motivation_perturbing_amp}, we relate 
the amplitude for the Weyl scalar $\Psi_4^{(1)}$ extrapolated to future null infinity to the Weyl scalar $\Psi_0^{(1)}$ evaluated at the black hole horizon via the NP equations.
This gives
\begin{align}
\label{eq:phys_amp_final}
&|\mathcal{A}| = |\Psi_0^{(HH)}(r_+)| \\
   &=C_{\text{lookback}}\left|\frac{C_{\text{to-horizon}}}{16 M^2r_+^3} \sqrt{\left|\frac{4D'}{\mathfrak{C}'}\right|^2 +\left|\frac{48 \omega M}{\mathfrak{C}'}\right|^2} \left(\frac{ |h_{lmn}\omega^2|}{2}\right)\right|
   .\nonumber
\end{align}
In the above, $D'$ and $\mathfrak{C}'$ are Starobinsky constants defined by Eqs. (2.25) and (2.28) in Ref.~\cite{Berens:2024czo}, and $C_{\text{to-horizon}}$ [defined in Eq.~\eqref{eq:to_horizon_psi0}] relates the solution at null infinity to one at the horizon.
The quantity $C_{\text{lookback}}$ refers to the time coordinate shift due to a slicing choice Eq.~\eqref{eq:look-back-time}. The uncertainty in this value relates to an uncertainty in the distance of the quasinormal mode peak from the black hole horizon.
We have also introduced the spheroidal components  
of the gravitational wave strain $h$ (times extraction radius $r$) of a quasinormal mode, indexed by an overtone number $n$, angular number $\ell$, and azimuthal angular number $m$,
\begin{align}
    \label{eq:hlmn}
    h_{\ell m n} = \int d\Omega \frac{h\ r}{M} \ _{-2}S_{\ell m n}^* ,
\end{align}
where the spheroidal harmonics are normalized with $\int_{-1}^{1} S S^*d\cos\theta =1$. 

Here we calculate the leading-order change in mass and spin due to the
absorption of the $\ell=2,\ m=2,\ n=0$ quasinormal mode, which we consider
since this mode has longest decay time of all quasinormal modes, and it has
been the most robustly measured of all ringdown modes \cite{LIGOScientific:2016lio,Ghosh:2021mrv}. To estimate the magnitude of such a quasinormal mode that is
relevant for the aftermath of a comparable mass binary black hole merger, we
consider the fits to ringdown modes that were computed from numerical
relativity simulations in Ref.~\cite{Giesler:2019uxc}. In that reference, the
authors analyzed a quasicircular binary black hole merger with mass ratio
$1.22$, which led to a remnant that had dimensionless spin $a/M = 0.692$, and
found a strain amplitude of $|h_{220}| \approx 0.41$ (converting to our
conventions) when fitting from the peak of the strain 
onward. Using Eq.~\eqref{eq:phys_amp_final}, this translates to
$|\mathcal{A}|\approx 0.013  (C_{\text{lookback}}/1.86)$,
where we estimate that $C_{\text{lookback}} = 1.86 \pm 0.9$ based on the light crossing time from the black hole light ring, with an uncertainty of $\delta r = ^{+M}_{-M/5}$ in Boyer-Lindquist radius (see Appendix~\ref{App:motivation_perturbing_amp} for further discussion). 

Integrating Eq.~\eqref{eq:area-change-psi0-mode-sol} from $v_1 = 0$ to $v_2 = \infty$ to find the change
in the black hole's area, and applying 
Eqs.~\eqref{eq:change_in_M_for_qnm_triangle} and \eqref{eq:change_in_a_for_qnm_triangle}, we can approximate the changes induced on this example black hole due to the partial absorption of its fundamental mode. We find, for remnant spin $a/M = 0.692$ and perturbing frequency $\omega_{220}(a/M)$,
\begin{align}
    \Delta M/ M &= \left(0.92\%\right) \times \left(\frac{C_{\text{lookback}}}{1.86}\right)^2\left(\frac{h_{220}}{0.408}\right)^2 ,\\
    \Delta a/a &= \left(6.34\%\right) \times \left(100\times \Delta M/ M\right)
    .
\end{align}
The change in the black hole
parameters scales with the square of the perturbing amplitude, and there is 
a factor of 2 uncertainty in $ C_{\text{lookback}}$ corresponding to a factor
of 4 uncertainty in $\Delta M$ and $\Delta a$. 
This change in background corresponds to a change in the quasinormal mode frequency of $\Delta \omega_{(220)}/\omega \approx 2.6\%$.  

\begin{figure}[h]
\centering
\includegraphics[width=0.5\textwidth]{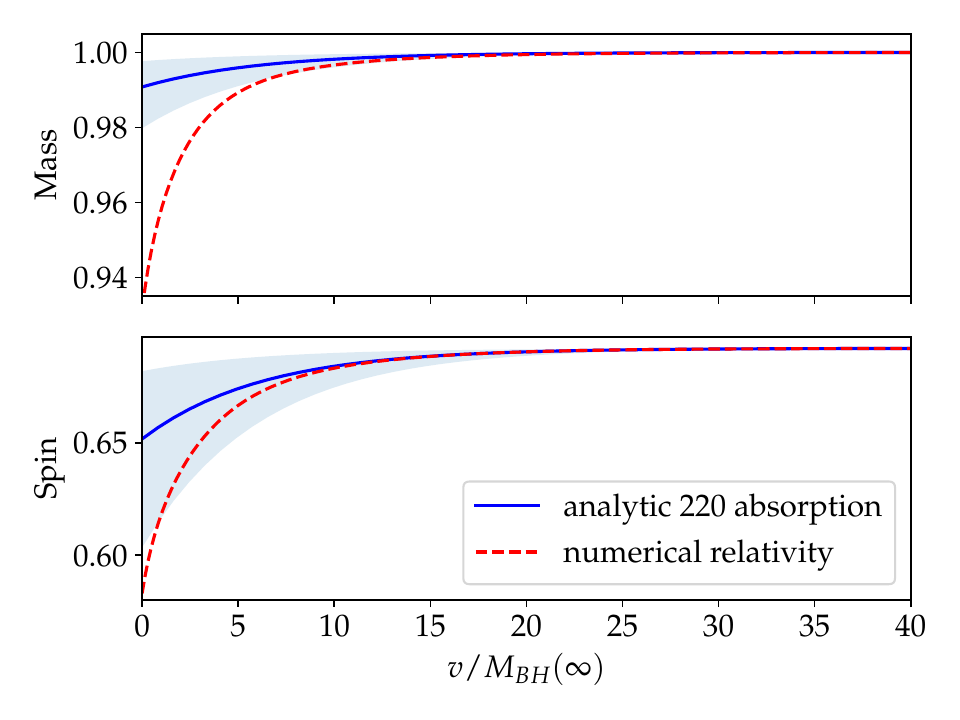}
\caption{Mass and spin of a binary black hole merger remnant according to numerical relativity calculations \cite{GieslerEvent}. Units are of the remnant mass, with ``spin'' corresponding to $\left[J(v)/M(v)\right] M(\infty)$. For the simulation, $v$ refers to the coordinate time and $v=0$ is when a common horizon is formed [the strain peaks at infinity at $v\approx7M(\infty)$]. We compare these numerical relativity results with our analytic calculation of the change in black hole parameters due to the absorption of a quasinormal mode with $\omega_{\ell m  n} = \omega_{220}$ using our estimate of its amplitude. 
We include our estimated error on the change in mass and spin due to the absorption of that mode. The high uncertainty in this calculation is due to an uncertainty in the lookback time (these error bars correspond to assuming a peak at the light ring with uncertainty of $1M$ farther from the black hole and $M/5$ closer to the black hole). Note that the time coordinate is different for the numerical relativity and analytic cases.  
    }
\label{fig:mass_spin_change_in_t}
\end{figure}

In Fig.~\ref{fig:mass_spin_change_in_t}, we show the result of this analytic calculation for $M(v)$ and $a(v)$. The timescale of the change in mass and spin is determined by the imaginary part of the perturbing mode. We compare this result to a full general relativity calculation for the same event that motivated our amplitude choice \cite{GieslerEvent}. (Recall, this black hole binary has a quasicircular inspiral with mass ratio $1.22$ and remnant dimensionless spin $a/M = 0.692$.) The data in \cite{GieslerEvent} include calculations of the apparent horizon (Christodoulou) mass and spin at each time step. These quantities are calculated using an approximate azimuthal Killing vector and are not expected to be exact. We plot the simulation time coordinates divided by the remnant mass, which will not correspond exactly to the time coordinate $v$ of the analytic calculation. In Ref.~\cite{Giesler:2019uxc}, the fit to quasinormal modes also found overtones with significant amplitude at early times, which could also contribute to the flux through the horizon; however, we only include the fundamental mode in our analytic calculation.
Because of these caveats, we only expect an approximate relation between the calculations from numerical relativity and our analytic estimate, even at late times. However, we find it reassuring that the plotted quantities are comparable.

It is informative to compare our total mass change calculation to the numerical relativity calculation postmerger. From the numerical relativity result, the black hole's mass changes by $\approx 6\%$ after forming a single horizon.
The change in spin is $\approx 17\%$ over the same period. Using our analytic calculation [considering only the $(\ell m n) = (220)$ mode], we found a mass change of $\approx1\%$, with a corresponding spin change of $\approx 6\%$. 
We can also look at the time dependence in the numerical relativity mass approximation. For time $t = (10-100)M$ (the last half a percent in mass change), the mass change is well described by $M = M(\infty)(1 - \exp{C v})$ with real constant $C$. 
Fitting the exponent $C$, we find that it matches the value expected from the absorption of the $(220)$ mode associated with the final black hole ($C  = 2 \mathfrak{I}\omega_{220}$) to within $\sim 0.5\%$.


\section{Methods}
\label{sec:methods}

Here we describe how we generate numerical waveforms using a modified Teukolsky evolution code that allows the black hole mass and spin to change with time in a prescribed way. We then detail how we identify quasinormal modes in the gravitational wave signal using a quasinormal mode filter \cite{Ma:2022wpv,Ma:2023vvr, Ma:2023cwe}. 
To check the robustness of our results, we then outline an another approach to fitting for quasinormal modes in the time domain using a least-squares fitting procedure. Once the mode content has been determined, mode amplitudes can be fit in the time domain. We describe our fitting template and how we apply it in this case.

We use the symbols $M_{\text{BH}}(t)$ and $a_{\text{BH}}(t)$ to denote the time-dependent black hole quantities, with $M_{\text{BH}}(\infty) = M$ and $a_{\text{BH}}(\infty) = a$.
We use $\mathcal{A}$ to denote the amplitude of $\Psi_4$ at null infinity, such that the perturbing quasinormal mode has the form
\begin{align}
    \Psi_4 = \mathcal{A}
   \times
   e^{im\varphi - i\omega t}
   \frac{R\left(r\right)}{R(\infty)} S \left(\vartheta\right)
    .
\end{align}


\subsection{Numerical evolution setup}
\label{sec:numerical_evolution_set_up}

We use a Teukolsky evolution code (henceforth TEC) based on the horizon-penetrating, hyperboloidally compactified (HPHC) coordinates described in Ref.s~\cite{Zhu:2023mzv,justin_ripley_2023_8302934} (for more discussion on these coordinates, see also Refs.~\cite{Ripley:2020xby,Ripley:2022ypi}).
Using TEC, we evolve the linearized Weyl scalar $\Psi_4^{(1)}$ by solving the Teukolsky equation.
As TEC evolves the Teukolsky equation in HPHC coordinates, the black hole horizon and future null infinity are both accessible in a finite amount of ``time'' from the interior. 
We construct quasinormal mode initial data using the methods of Ref.~\cite{Ripley:2022ypi}. 

For this study, we have made several minor modifications to the version of TEC described in Ref.~\cite{Zhu:2023mzv}. 
We have adapted the code to recalculate evolution operators at each time substep using a parametrized black hole mass and spin.
We also made minor changes to both the initial data and evolution codes to allow for arbitrary grid choices (instead of always using a grid corresponding to the fixed physical background), which we use to construct initial data and evolve on a grid where the inner boundary is always set to $r=M$ in Boyer-Lindquist coordinates. 
Using a fixed grid makes it more straightforward to compare the results of simulations with different backgrounds. 
In what follows, $t$ corresponds to the hyperboloidal time coordinate defined in Ref.~\cite{Zhu:2023mzv}.

To analyze the results of our numerical simulations, we extract $\Psi_4(t,\theta,\varphi)$ from a sphere at future null infinity, and project out its spin-weighted spherical harmonic components $\Psi_4^{\ell m}(t)$. 
We note that each component $\Psi_4^{\ell m}(t)$ does not generally describe an individual quasinormal mode, as for nonspinning black holes the angular dependence of the quasinormal modes are described by ``spin-weighted spheroidal harmonics'' \cite{Teukolsky:1973ha}, not spin-weighted spherical harmonics,
\begin{align}
    _{s}S_{\ell m}(\theta,\phi ; a\omega) &= 
    \sum_{\ell'=\ell_{\text{min}}(s,m)}^{\ell_{\text{max}}} c_{\ell' \ell m}\ _{s}Y_{\ell' m}(\theta,\phi)\\
    &c_{\ell' \ell m}(a>0) \neq 0
    .
\end{align}
We calculate these factors of $c_{\ell' \ell m}$ using Ref.~\cite{Stein:2019mop} and use them to weight amplitudes extracted from a particular spherical harmonic projection of the data.

When we begin a numerical evolution, we expect the first few $M$ of data in the time domain to be dominated by the initial data in the wave zone, and to be only weakly affected by the background. We want to choose a reference time where the data at null infinity start to be strongly dependent on the background. 
To estimate this, we calculate the propagation time from the black hole light ring to null infinity and to the horizon---see Appendix \ref{App:propagation_times}. 
For a background black hole with spin $a/M = 0.7$, we calculate a delay time of $9M$. For this case, we analyze data with $t>9M$, where we are sure that the signal at null infinity is from initial data that were evolved from the near black hole region, and not from the wave zone. Choosing a later reference time is more conservative, giving a smaller measurement for excited amplitudes relative to the initial perturbation. This is because the chosen perturbation is longer lived than the excited modes, and so their ratio decreases with time. The uncertainty in lookback time $\delta t$ introduces an uncertainty in any amplitude fits, with $\delta \left(A_{\ell m n}/\mathcal{A}\right) = e^{(\mathfrak{I}\omega_{\ell m n}-\mathfrak{I}\omega_{220})\delta t}$.

\subsection{Fitting quasinormal modes}
\label{sec:fittinq_qnm}

We adopt three steps for quasinormal mode fitting. 
To summarize, first we use quasinormal mode filters \cite{Ma:2022wpv,Ma:2023vvr, Ma:2023cwe} to select candidate modes. 
This filtering method enables the identification and subtraction of modes without requiring corresponding amplitude information, thereby eliminating the need for fitting and avoiding the risk of overfitting. 
Second, we implement free frequency fits for the candidate modes with varied starting times. 
We identify a mode when the fit converges to the targeted quasinormal frequency. 
In some cases, we perform the fit after filtering out other candidate modes, since it reduces the number of fitting parameters and improves fit performance. 
Once all the quasinormal modes are identified, we finally fit for their complex amplitudes while keeping frequencies fixed. 
Below, we provide more details about our procedure.

\begin{figure}[h]
\centering
\includegraphics[width=0.5\textwidth]{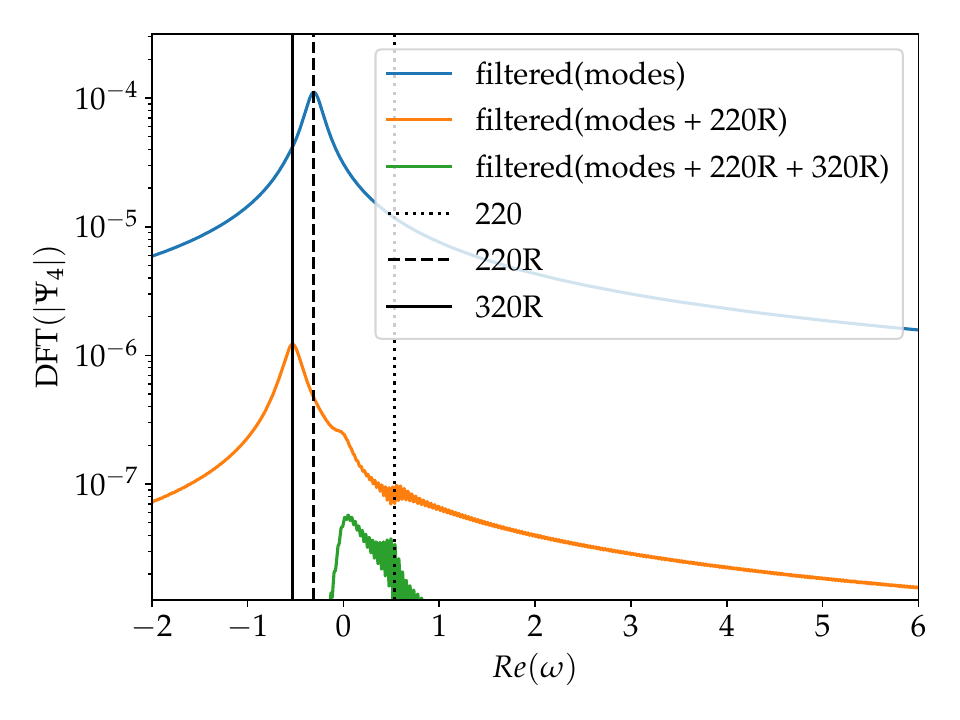}
    \caption{Fourier transform of a gravitational wave signal on a fixed background and remnant spin $a/M=0.7$. Here we see that there is some evidence for both the $220_R$ and the $320_R$ frequency in the data. Here ``modes'' corresponds to: 
    $(22n)$ with $n=0,\ldots,7$; $(320)$; $(221)_R$; and $(222)_R$.
    }
\label{fig:mode_id_example_FFT}
\end{figure}

In Ref.~\cite{Ma:2022wpv}, the authors introduced a frequency-domain filter for a quasinormal mode $\omega_{\ell mn}$,
\begin{align}
    \label{eq:qnm_filter}
    F_\omega=\frac{\omega-\omega_{\ell mn}}{\omega-\omega_{\ell mn}^*}.
\end{align}
After filtering out the dominant mode, they demonstrated the presence of oscillations induced by subdominant modes in time-domain waveforms, thereby aiding in the identification of these modes. 
Here, we extend this analysis by examining the Fourier transformation of filtered waveforms. 
In Fig.~\ref{fig:mode_id_example_FFT}, we show how to use the Fourier spectrum to identify modes using 
an example consisting of a fixed background and remnant spin $a/M=0.7$, but where the initial data consist of
a fundamental quasinormal mode with respect to a black hole with slightly different parameters. 
After applying a discrete Fourier transform (DFT), several filters (listed in the figure) are applied to the data, resulting in
the blue curve.
There is a prominent peak at the real frequency of the $(220)_R$ retrograde mode (the negative center frequency indicates it is a retrograde mode). Retrograde modes are modes that travel in the opposite direction around the black hole relative to its spin. Here, we use a subscript $R$ to indicate that a mode is retrograde. They have frequency
\begin{align}
    \omega_{\ell m n R} = -\mathfrak{R}\omega_{\ell -m n} + i\mathfrak{I}\omega_{\ell -m n} \label{eq:def_retrograde}
    .
\end{align}
The orange curve shows the result of the DFT after applying another filter to remove the $(220)_R$ mode. Clearly, this filter removes the peak at $(220)_R$. It also reveals a smaller new bump in the negative frequency band, corresponding to $(\ell,m,n)=(320)_{R}$. The green curve shows the result after filtering out $(\ell,m,n)=(320)_{R}$, removing the peak at $(\ell,m,n)=(320)_{R}$. In this example, we have identified two modes potentially present in the signal by examining DFTs of the filtered waveform. We repeat this procedure iteratively to determine all the candidate modes.

\begin{figure}%
    \centering
    \subfloat[\centering Free frequency fits in complex plane]{{\includegraphics[width=8cm]{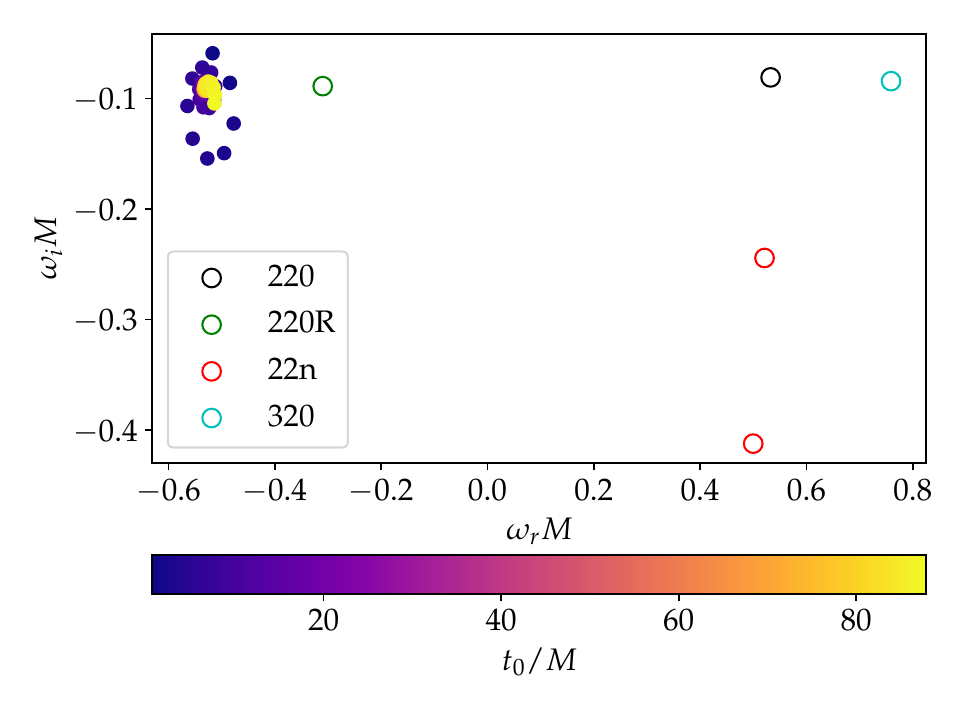} }}%
    \qquad
    \subfloat[\centering $\left|\omega_{\text{fit}}-\omega_{320R}\right|M$]{{\includegraphics[width=8cm]{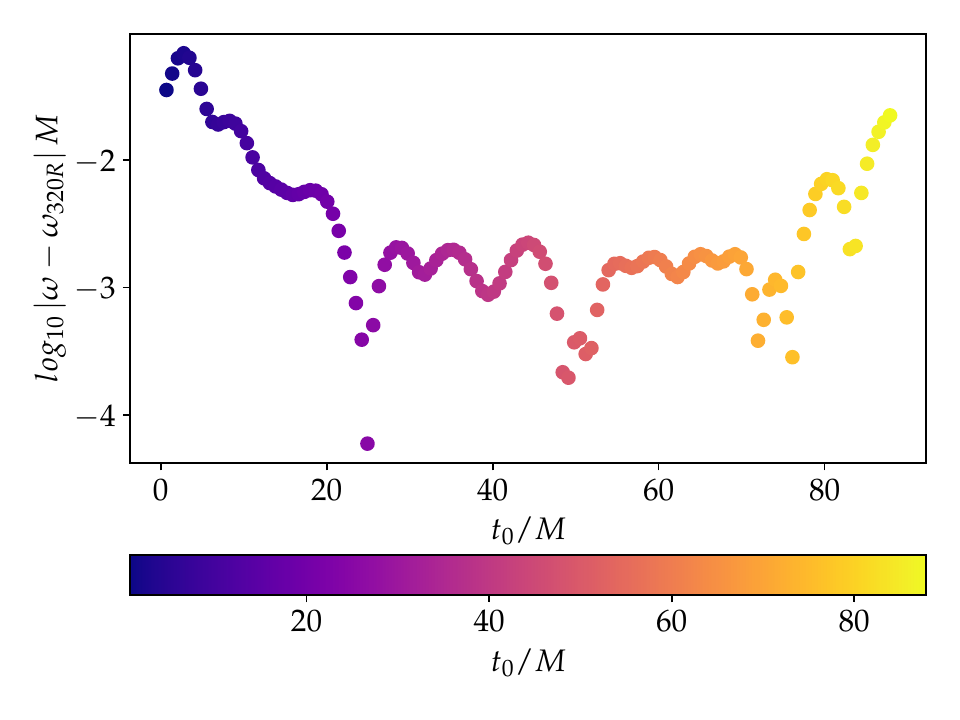} }}%
    \caption{We perform a free frequency fit on $\Psi_4$ after filtering out previously identified modes except for $(\ell, m, n) = (320)_R$. Fits are performed for data in the window $t\in(t_0,100M)$. In the complex plane, the free frequency fits are clustered around the physical value of $\omega_{320R}$. In  (b), we plot the absolute difference between $\omega_{\text{fit}}=\omega_r+i\omega_i $ and the physical $\omega_{320R}$. From $\left|\omega_{\text{fit}}-\omega_{320R}\right|M$, we can see that in the first $\sim 50M$ the fit frequencies converge to the value of $\omega_{320R}$. At later times, the frequency fits diverge from this as the fitting window becomes small. From the combination of these results and Fig.~\ref{fig:mode_id_example_FFT}, we conclude that $(320)_R$ is present in the data. 
    }%
    \label{fig:mode_id_example}%
\end{figure}

Once we have identified candidate modes using the filter, we use a free frequency fit to confirm the presence of that mode. Our free frequency fits use a model where a single free complex exponential is fit to data in different time windows,
\begin{align}
    \label{eq:free_freq_model}
    \psi^{\text{free freq}} =  A \exp{\left[-i(\omega_r+i\omega_i) t\right]} ,
\end{align}
where the free parameters are $A\in \Bbb{C}$ and $\omega_r, \omega_i \in \Bbb{R}$.
These fits are useful for confirming the presence of a mode in data after its real frequency is identified using the filter. In Fig.~\ref{fig:mode_id_example}, a free frequency fit is performed on the data after filtering out all of the modes in the list of modes already identified, except for $(\ell, m, n) = (320)_R$. We see a clustering of points around the value of the remnant $(320)_R$ mode. We conclude that this mode is physically present. 

After determining the mode content, we perform fixed-frequency fits. We use the model
\begin{align}
    \label{eq:fixed-freq-model}
    \psi^{\text{fixed-freq}} = \sum_{1}^{N} A_N \exp{\left[-i \omega_N t\right]},
\end{align}
where the free parameters are $A_N\in \Bbb{C}$, and $\omega_N$ are the fixed frequencies of quasinormal modes calculated using linear perturbation theory. So for $N$ modes, this model has $2N$ real free parameters.
In order to determine the physical amplitude and extract it, we vary the time window of data considered in the fit, as in Ref.~\cite{Baibhav:2023clw}. We introduce an arbitrary reference time to describe the fitted amplitudes in a time-independent way. The time-independent amplitude is defined relative to the amplitude fit at $t=t_0$,
\begin{align}\label{eq:amp_ref_time}
    \left|A^{t_{\text{ref}}}_{N}\right| = \left|A^{\text{fit}}_{N}(t_0)\right|\exp{\left(-\mathfrak{I}\omega_{N}(t_0-t_{\text{ref}})\right)}
    .
\end{align}
Then a region where the fitted amplitude has the expected decay rate corresponds to a region where the time-independent fitted amplitude is constant. We vary the start of the time window and check if the time-independent fitted amplitude is constant in some region. This region would correspond to a time window where the mode is resolved, and where the signal is well described by a superposition of quasinormal modes with fixed amplitudes. 
We fit mode amplitudes in the time domain using a least-squares algorithm\footnote{We use the Levenberg-Marquardt method as implemented in \texttt{scipy.optimize.least\_squares}. However, since the amplitude is a linear parameter, it can be solved exactly, without optimization, by inverting a matrix representation of the Fourier transform of the model.} to minimize the difference between the fit model and the signal. 

In general, we apply filters to the data before fitting mode amplitudes. Note that applying the filter to a mode will affect the amplitude of that mode. Applying a filter $F_{\omega}$ to a mode with frequency $\omega'$ gives
\begin{align}
    F_\omega &\cdot \exp{(i \omega' t + \phi)} 
    \\&=
    \begin{cases}
    0& \text{if } \omega = \omega',\\
    A^{\text{filter}}(\omega,\omega')\exp{(i \omega' t + \phi')}  & \text{otherwise} 
    .\nonumber
\end{cases} 
\end{align}
This amplitude change has the form
\begin{align}
    A^{\text{filter}}(\omega,\omega') = \frac{\omega' - \omega}{\omega'-\omega^*}
    ,
\end{align}
where $\omega^*$ is the complex conjugate. Applying multiple filters results in an amplitude change given by the product of these suppression factors. We use Ref.~\cite{Stein:2019mop} to calculate quasinormal mode frequencies to high precision as input to $A^{\text{filter}}(\omega,\omega')$. Then we divide any amplitudes fit to filtered data by this suppression factor to recover the amplitude in the unfiltered signal. We test our numerical computation of the filter factor by comparing the amplitude change of a filtered pure mode to the product of $O(10)$ suppression factors. We find agreement to the level of $0.01\%$.
We weight extracted amplitudes both by their suppression by the filter and by the spherical-spheroidal mixing component, so the final amplitude results are with respect to the spheroidal basis.

\section{Results}
\label{sec:Results}
In Sec.~\ref{sec:quasi-normal_mode_absorption},
we determined to leading order how we expect the mass and spin of the black hole to change with time in the presence of a particular quasinormal mode. Here we present the results of our numerical experiments, probing how a changing black hole background affects a single perturbing quasinormal mode. The change in background is second-order in perturbing amplitude, so the change in the perturbation is expected to be third order,
\begin{align}
    \Psi_4 = \Psi_4^{\text{initial}}\left[ O(\mathcal{A})\right] + \Psi_4^{\text{excited}}\left[ O(\mathcal{A}^3)\right]
    .
    \label{eqn:psi4_perturb}
\end{align}

\begin{figure}%
    \centering
    \subfloat[\centering Real part of $\Psi_4$]{{\includegraphics[width=8cm]{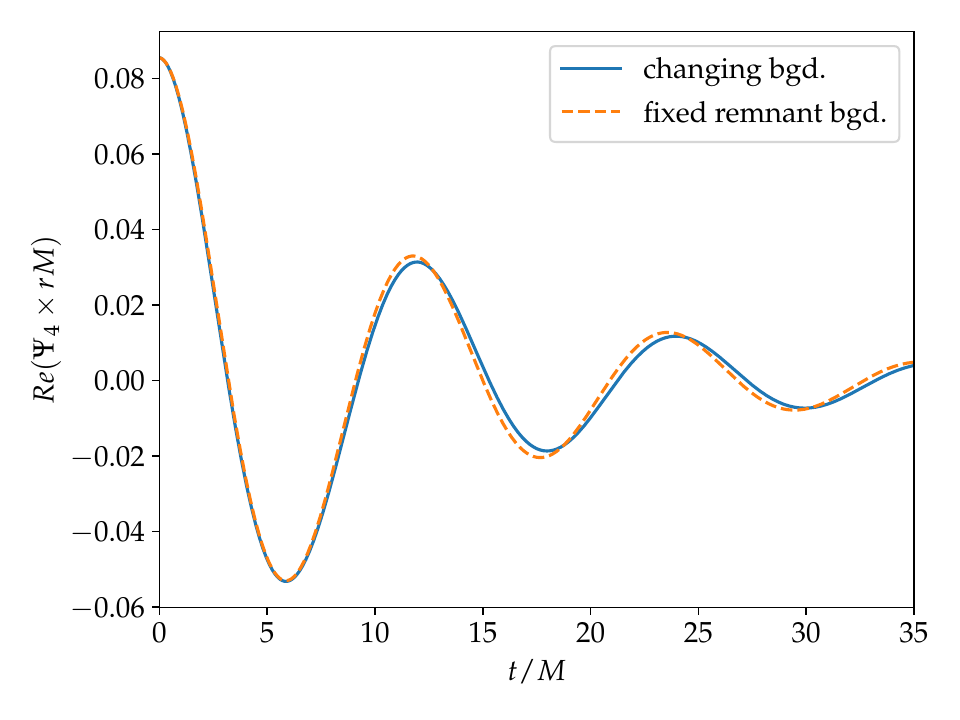} }}%
    \qquad
    \subfloat[\centering Difference in phase of the two signals]{{\includegraphics[width=8cm]{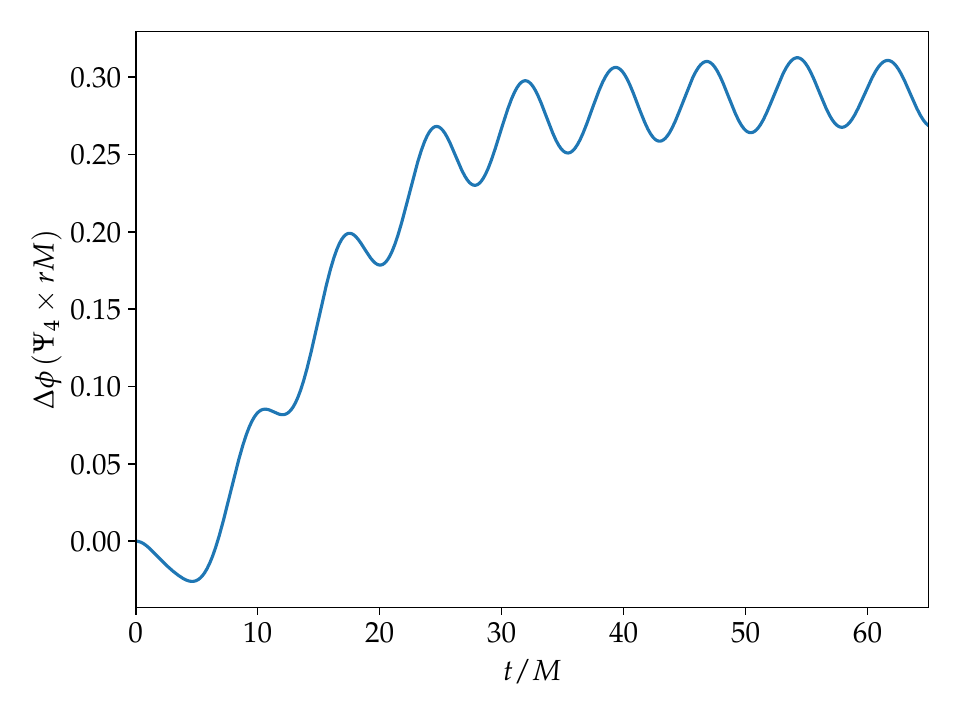} }}%
    \caption{(a) The real part of $\Psi_4$ at null infinity for a changing background case with $\Delta M/M = 4.7\%$ compared to a fixed black hole background. Both cases have identical initial perturbation and remnant black hole parameters $M_{\text{BH}}(\infty) = M$, $a_{\text{BH}}(\infty) = a=0.7M$. The change in the signal is on the level of a few percent. (b) The phase difference between the two signals. The phase difference asymptotes to a constant at late times.
    }%
    \label{fig:Repsi_change}%
\end{figure}

In this section, we study the change in a pure quasinormal mode perturbation
due to the change it induces on the black hole. Given an initial perturbation,
we can calculate the resulting background change. Using our flux calculation,
we determine an initial black hole that, in the presence of the fundamental
$(\ell=2,m=2,n=0)$ quasinormal mode of that remnant, will change mass and spin
to have exactly the properties of that remnant. We use the remnant quasinormal
mode as initial data on a background that changes by $\Delta M$ and $\Delta a$
from the determined initial black hole to the remnant according to our flux
calculation. So we set the background to have mass and spin:
\begin{align}
    \label{eq:adiabatically_changing_background_M}
    M_{\text{BH}}(t) &= M- \Delta M \exp{\left[2\mathfrak{I}\omega_{220} t\right]}, \\
    \label{eq:adiabatically_changing_background_a}
    a_{\text{BH}}(t) &=a- \Delta a \exp{\left[2\mathfrak{I}\omega_{220} t\right]} .
\end{align}
We compare this case with the same initial data dissipating on the fixed remnant background with mass $M$ and spin $a$, i.e. a
pure quasinormal mode, in Fig.~\ref{fig:Repsi_change}.
As we will detail below, the perturbation to the quasinormal mode from the changing background can be characterized by both a change to the amplitude and phase of the original, fundamental quasinormal mode, as well as the excitation of additional quasinormal modes.

We evolve the linear Teukolsky equations for $\Psi_4$ on a black hole
background evolving according to
Eqs.~\eqref{eq:adiabatically_changing_background_M} and \eqref{eq:adiabatically_changing_background_a},
with some small $\Delta M$ and $\Delta a\propto \Delta M$.  When computing the
resulting shift in $\Psi_4$ with respect to the fixed
background case (i.e. a pure quasinormal mode), to leading order we will get
a linear-order dependence on $\Delta M$, and an additional linear dependence on
$\mathcal{A}$ (the initial amplitude of $\Psi_4$), which combine to give the
cubic dependence in Eq.~\eqref{eqn:psi4_perturb}. This means that our results
can be rescaled to different values of $\Delta M$, e.g., $\Delta A_{\ell m
n}/{\mathcal{A}} \propto \Delta M$ for an excited mode amplitude, provided
this quantity is sufficiently small.

\begin{figure}[h]
\centering
    \subfloat[\label{fig:signal_change_a}\centering Dominance of linear contribution with $a/M = 0.7$]{{\includegraphics[width=8cm]{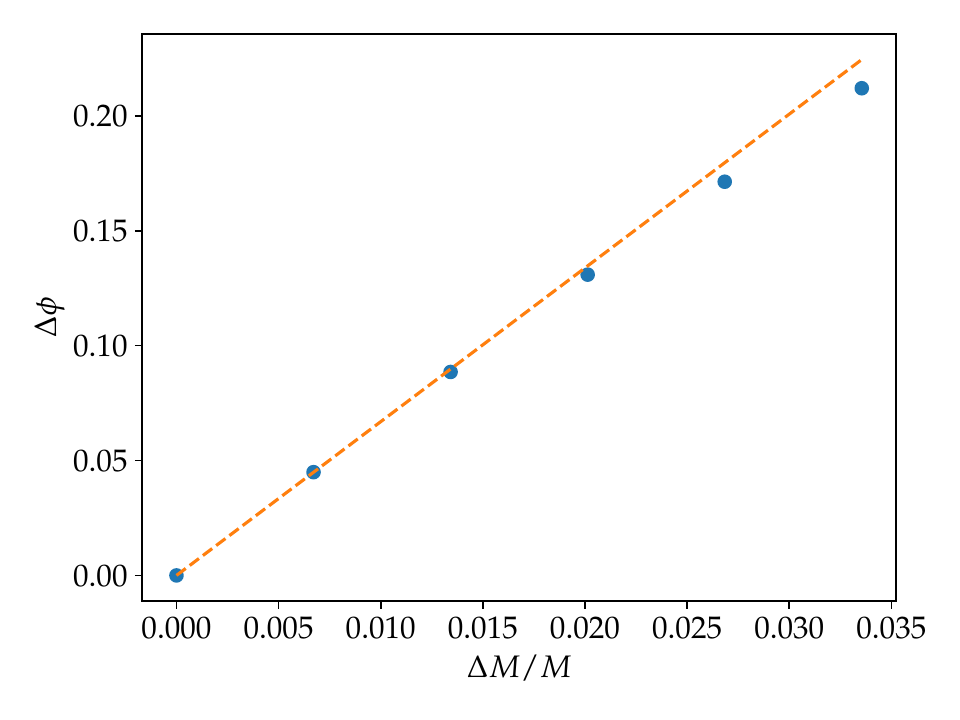} }}%
    \qquad
    \subfloat[\label{fig:signal_change_b}\centering Dependence of $\Delta \phi$ on spin with $\Delta M/M = 1\%$]{{\includegraphics[width=8cm]{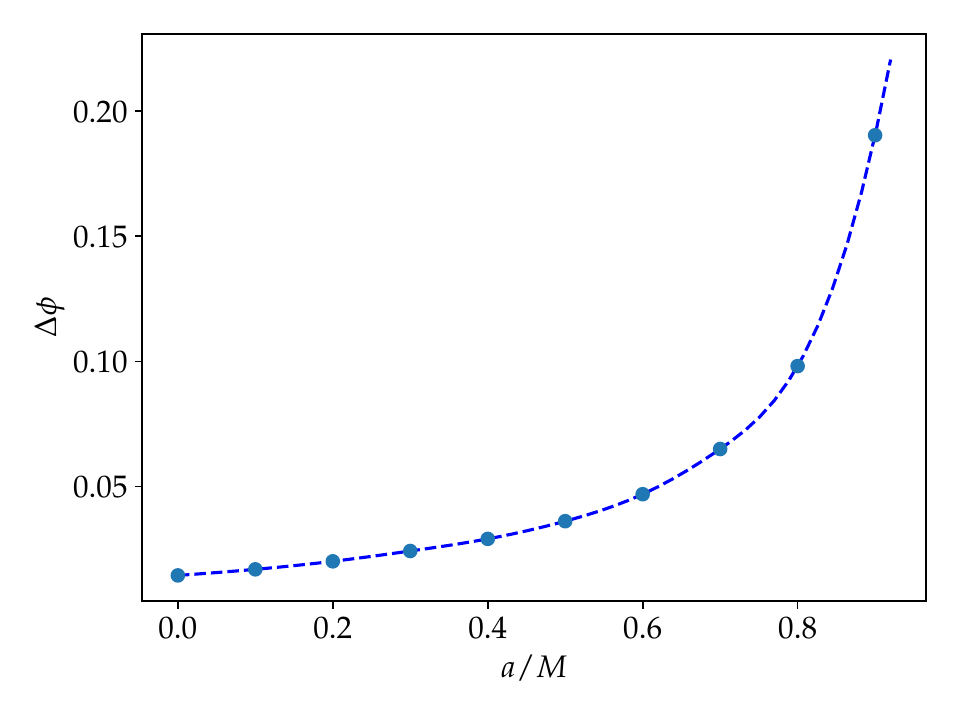} }}
    \caption{Difference of signal due to the third-order horizon flux effect. Comparing a signal with changing background $\Psi_4(\Delta M)$ to a signal from a fixed background $\Psi_4(0)$. We calculate 
    the accumulated phase offset due to the changing background by calculating $\Delta \phi = \phi(\Delta M)-\phi(\Delta M = 0)$, which asymptotes to a constant
    at later times ($t\geq 40M$, see Fig.~\ref{fig:phase_change_diff_dMs}). In (a), the phase offset is calculated for several mass changes. The dashed line is a linear fit to the first two points. We determine that the relative change in signal is dominated by a term linear in the mass and spin change for the range plotted, with a $6\%$ deviation for $\Delta M/M = 3.4\%$. (b) The dependence of the phase offset on spin. In this case, the dashed line is an interpolation of the points.
    }
\label{fig:signal_change}
\end{figure}

Technically, the coefficients in the Teukolsky equation depend nonlinearly on the black hole mass and spin,
but we demonstrate that we recover the expected scaling for the values within the range of interest
by simulating several systems with
varying $\Delta M$ and showing the resulting change in gravitational wave signal
in Fig.~\ref{fig:signal_change_a}. Here we use the accumulated phase offset
$\Delta \phi$ as a relatively clean measure of the change in signal.
$\Delta\phi$ is the phase of $\Delta\Psi_4$, $\Delta \phi = \phi(\Delta
M)-\phi(\Delta M = 0)$, which asymptotes to a constant at late times. As
shown, for these choices of $\mathcal{A}$, the leading-order term dominates, and
we have $\Delta\phi\propto \Delta M$ within $6\%$ for $\Delta M/M
\lesssim 3.5\%$. 

In Fig.~\ref{fig:signal_change_b}, we show the dependence of
the accumulated phase on spin. The phase offset increases significantly for
higher spin, which can be explained by the quasinormal mode frequency shifts.
In general, for some fixed change in dimensionless spin, the change in quasinormal
mode frequency is larger for larger initial spin i.e., $d \omega_{220}/ d(a/M)$ increases
monotonically with $a/M$. 

This change in gravitational wave signal can be characterized by a change in the perturbing mode amplitude $\Delta A_{2 2 0}$, along with the excitation of a superposition of extra modes, with excited amplitudes $A_{\ell m n}$. 
To determine which modes are excited, we use a combination of frequency-space filtering and free frequency fits in the time domain, according to the methods described in Sec.~\ref{sec:fittinq_qnm}. Once the relevant modes are determined, we can fit their amplitudes in the time domain. 

\begin{figure}[h]
\centering
\includegraphics[width=0.5\textwidth]{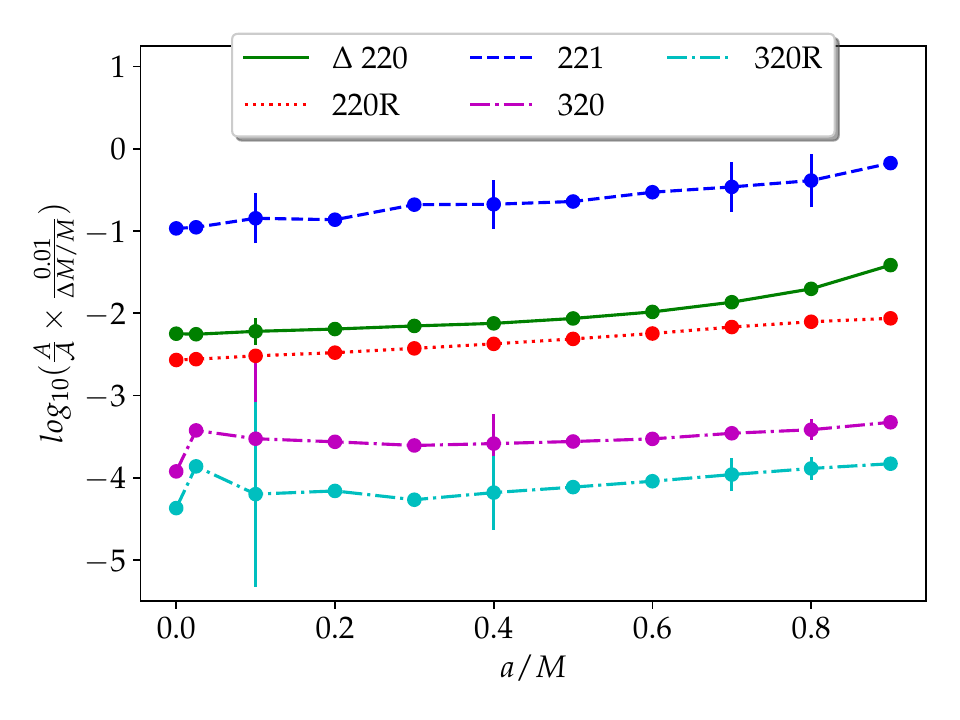}
    \caption{Amplitudes of excited modes $A_{\ell m n}$ as function of spin.
    These amplitudes are fit from the $(\ell m) = (22)$ spherical harmonic
    projection of the data. Higher $\ell$ modes are less excited for lower spin
    and become badly resolved below $a/M = 0.3$---in the $a/M = 0$ case the
    mixing coefficient between the $\ell = 3$ spheroidal harmonic and the
    $\ell = 2$ spherical harmonic becomes zero. In this case we fit the $\ell =
    3$ result from the $\ell m = 32$ spherical projection of the data. These
    amplitudes are all calculated for a mass change $\Delta M = 1\%$ and a spin
    change consistent with that. The error bars are calculated for some of the
    spin cases, they include the numerical error and error due to uncertainty
    in lookback time. The numerical error is determined by comparing the values
    extracted from simulations with two resolutions (see Appendix~\ref{sec:num_convergence}). The
    error due to an uncertainty in lookback time $\delta t$ is a factor of
    $\exp\left[i(\omega_{\ell m n}-\omega_{220})\delta t\right]$.}
\label{fig:spin_dep_amps}
\end{figure}

\begin{figure}[h]
\centering
\includegraphics[width=0.5\textwidth]{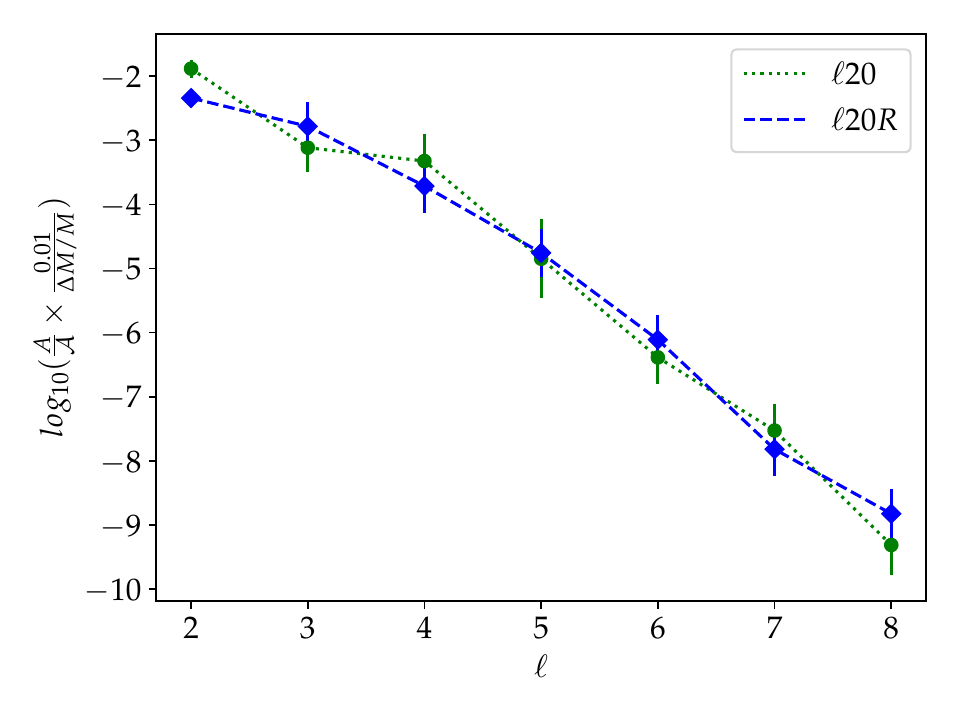}
\caption{Amplitudes of excited modes $A_{\ell 2 0}$ for a remnant with spin $a/M=0.7$ and perturbing amplitude such that $\Delta M /M= 1\%$. These modes were fit from spherical modes $\psi_{\ell',m}$ with $\ell = \ell'$, $m=2$. The error bars are the numerical error, determined by comparing the values extracted from simulations with two resolutions and assuming first order convergence in $A/\mathcal{A}$ (see Appendix \ref{sec:num_convergence}).}
\label{fig:higher_l_modes}
\end{figure}

When analyzing $\Psi_4$ at null infinity, we find retrograde modes,
higher $\ell$ modes and, potentially, overtones. We recall from the last section
that retrograde modes are modes that travel in the opposite direction around
the black hole to its spin, see Eq.~\eqref{eq:def_retrograde}. We perform a
mode analysis for several spins, fitting amplitudes in each case as outlined in
Sec \ref{sec:fittinq_qnm}. The results of fitting the $\ell= m = 2$ spherical
harmonic are shown in Fig.~\ref{fig:spin_dep_amps}. There is some ambiguity in
the overtone fit, where the overtone amplitude appears to increase
at early times. This may be due to the presence of nonmode
content with frequency close to the $(221)$ overtone. This feature will be
discussed in detail in Sec.~\ref{sec:nonmode_content}. Here we provide our best
fit for the overtone amplitude with the caveat that the fit is not exact. Apart
from the overtone, the amplitude of the other modes can be confidently measured without any
significant dependence on the extraction time, as shown below. We also
perform a mode analysis on higher-order spherical harmonics, with $\ell \neq
2$. The result of this analysis is shown in Fig.~\ref{fig:higher_l_modes}.

In Appendix \ref{sec:Physical_case}, we compare the results above to those obtained using a variation on the initial data where we use a quasinormal mode of the initial black hole, rather than the final black hole, as an initial perturbation. We find that the same modes are excited in both cases, though quantitatively, the excitation of additional modes is larger when the initial data correspond to a quasinormal mode of the initial black hole. This means the results shown here are conservative in terms of estimating the size of this effect.

\section{NonMode Content}
\label{sec:nonmode_content}
Here we discuss the ambiguity in the overtone amplitude extraction and present some evidence for nonmode content in the data. This nonmode content is only present in cases with a background that evolves according to our flux calculation. We discuss the possible effect of a changing frequency in the data and how this kind of nonmode content can manifest.

\subsection{Ambiguity in overtone extraction}
\label{sec:ResultsII}

\begin{figure}[h]
\centering
\includegraphics[width=0.5\textwidth]{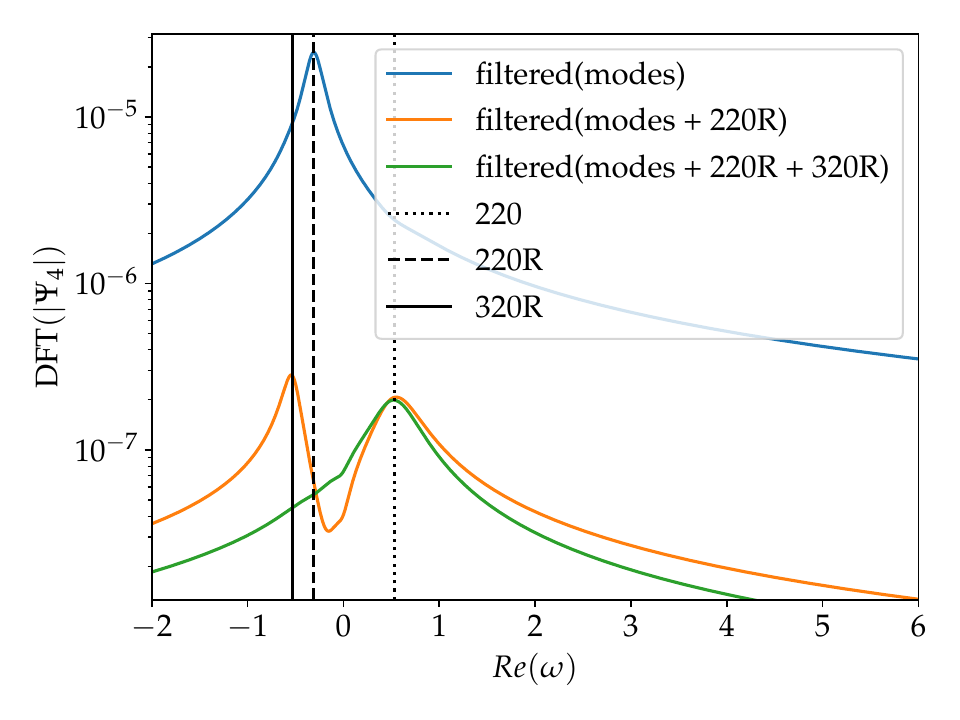}
    \caption{The results of performing a DFT on the gravitational wave signal with an evolving background and remnant spin $a/M=0.7$. We perform the same analysis as in Sec.~\ref{sec:fittinq_qnm}, Fig.~\ref{fig:mode_id_example_FFT}. Again, we see that there is some evidence for retrograde frequencies in the data. Recall that modes corresponds to the list: $(220)$, $(320)$, $(221)$, $(222)$, $(223)$, $(224)$, $(225)$, $(226)$, $(227)$, $(221)_R$, $(222)_R$. 
    Here, even after $(220)$ and seven of its overtones are filtered out, there is a peak at $\mathfrak{R}\omega_{220}$. Filtering out more overtones does not seem to remove this peak. 
    }
\label{fig:physical_FFT}
\end{figure}

When using a changing background, we find some evidence for an additional component to the gravitational wave signal that is not 
described by a sum of quasinormal modes with fixed frequency.
We first see this nonmode content in the frequency domain, using the filter analysis described in Sec.~\ref{sec:fittinq_qnm}. An example Fourier analysis is shown in Fig.~\ref{fig:physical_FFT}. In contrast to Fig.~\ref{fig:mode_id_example_FFT} above, where a fixed background was used, the spectrum in Fig.~\ref{fig:physical_FFT} indicates a component to the signal with similar real frequency to the $(220)$ mode, even though this mode and seven of its overtones have been removed by filters.  

\begin{figure}[h]
\centering
\includegraphics[width=0.5\textwidth]{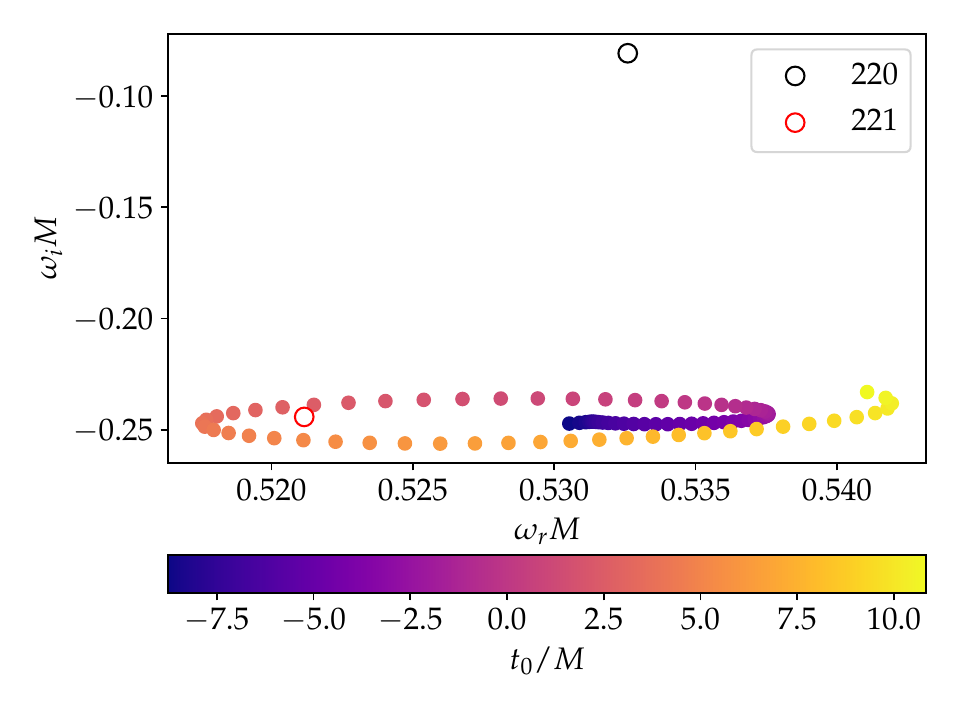}
\caption{A free frequency fit is performed after filtering out all identified modes, including seven overtones of $\ell=2$, $m=2$. The remnant is a black hole with dimensionless spin $a/M=0.7$, and the background evolves according to the absorption of a $(220)$ mode, with $\Delta M/M = 0.67\%$. The $x$ axis here stretches back before the lookback time (defined as $t=0$, see Sec.~\ref{sec:numerical_evolution_set_up}) to show the behavior from the start of the simulation. The initial perturbation is the $(220)$ mode of the remnant black hole. Note that this fit identifies a frequency with $\omega \approx \mathfrak{R}\omega_{220}+i\mathfrak{I}\omega_{221} \approx \mathfrak{R}\omega_{220}+i 3\mathfrak{I}\omega_{220}$ at early times, where these two complex frequencies are indicated, respectively, by the black and red circles.}
\label{fig:filtered_allmodes_changingbkgd}
\end{figure}

We use free frequency fits to better understand this contribution to the
signal. For an example with remnant spin $a/M = 0.7$, all identified modes are
filtered from the signal, and a free frequency fit is performed. This fit is
shown in Fig.~\ref{fig:filtered_allmodes_changingbkgd}. We confirm the result
from the frequency-domain analysis that there is some nonmode content with real frequency close to
the $(22n)$ modes. We see that this part of the signal has a similar decay rate
to the first overtone, or 3 times the decay rate of the fundamental mode. At very early times, the fit is close to $\omega =
\mathfrak{R}\omega_{220}+i\mathfrak{I}\omega_{221}\approx \mathfrak{R}\omega_{220}+i 3\mathfrak{I}\omega_{220}$. Since this frequency is
close to the first overtone it will be highly suppressed by the filter relative
to any numerical noise or other mode features. This means that this feature is
more significant, and better resolved at later times, in data with fewer
overtone filters (in particular, without filtering the first overtone).  We gain some
insight into a possible source for nonmode content with a similar decay
rate to the first overtone from a spherically symmetric model below, in
Sec.~\ref{sec:changing_freq}, and conclude that this may be due to a changing
frequency contribution. 

\begin{figure}[h]
\centering
    \subfloat[\centering Remnant quasinormal mode perturbation]{{\includegraphics[width=8cm]{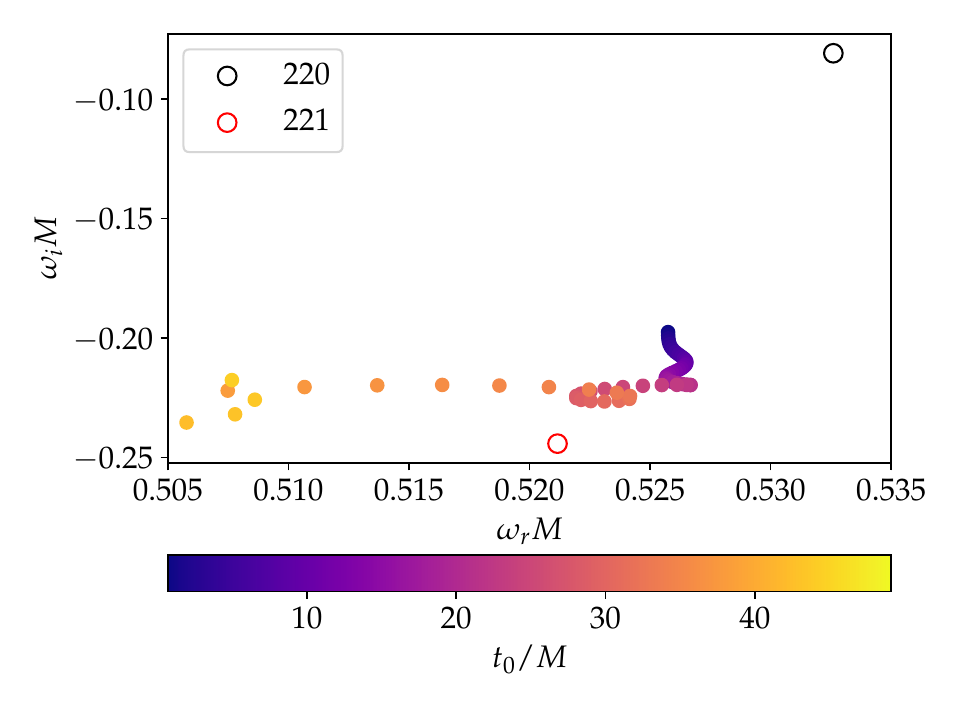} }\label{fig:free_freq_overtonea}}%
    \qquad
    \subfloat[\centering Initial quasinormal mode perturbation]{{\includegraphics[width=8cm]{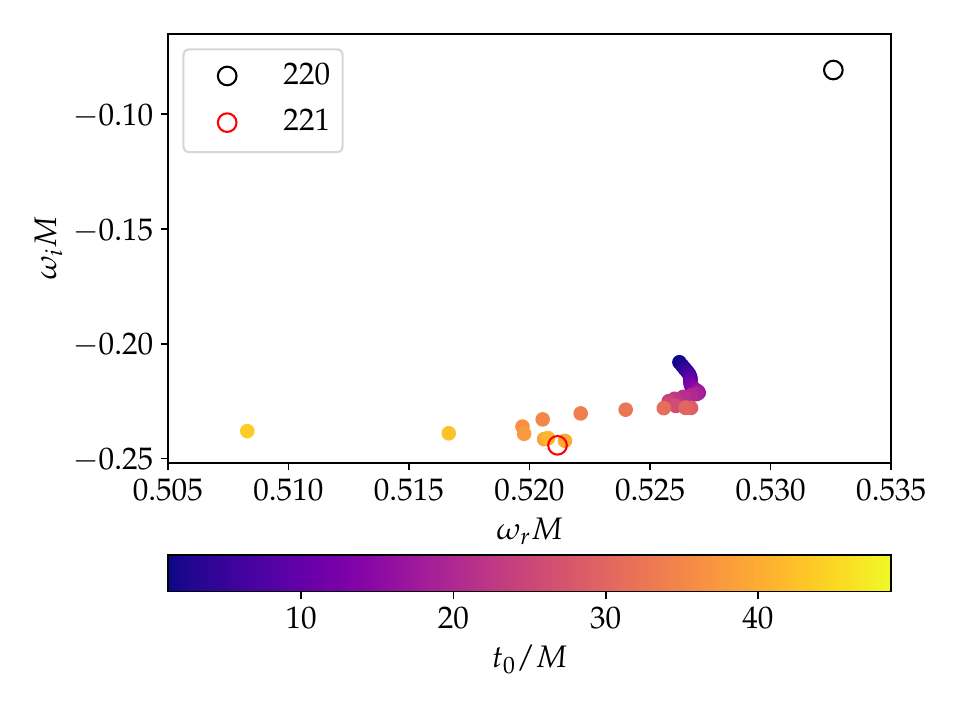} \label{fig:free_freq_overtoneb}}}
\caption{A free frequency fit is performed after filtering out modes except for the first three overtones $(221)$, $(222)$, $(223)$. The remnant black hole has dimensionless spin $a/M=0.7$, and the background evolves according to the absorption of a $(220)$ mode, with $\Delta M/M = 0.67\%$. These two plots correspond to two different initial data cases. (a) An initial perturbation equal to the $(220)$ quasinormal mode of the ``remnant'' black hole. (b) An initial perturbation equal to the $(220)$ quasinormal mode of the ``initial'' black hole i.e., the black hole with mass $M-\Delta M$ and spin $a-\Delta a$. In this case, the quickly decaying mode is resolved at later times and is shown to approach the overtone frequency with more confidence. We discuss this initial data case further in Appendix \ref{sec:Physical_case}.}
\label{fig:free_freq_overtone}
\end{figure}

Still considering the same example, in Fig.~\ref{fig:free_freq_overtonea} we filter the same modes as in Fig.~\ref{fig:filtered_allmodes_changingbkgd}, except for the $(22n)$ overtones with $n\in\{1,2,3\}$, and perform a free frequency fit. We see a shift in the fit decay rate (the imaginary part of the frequency), indicating that this component has some overlap with the overtones or is otherwise affected by the overtone filters. It seems that at early times the free frequency is converging to the first overtone, but that the overtone stops being resolved before the frequency reaches a constant value in time. We see frequencies with similar real part of the first overtone and the fundamental mode, with a decay time slower than the overtone. At $t=10M$ the free frequency fit gives a result close to the real frequency of the fundamental mode and with decay rate $\approx 10\%$ slower than the first overtone. 

For comparison, in Fig.~\ref{fig:free_freq_overtoneb} we consider a case where
the initial data are the $(220)$ quasinormal mode of the initial black hole, instead
of the final black hole. In this case we more confidently identify an overtone
amplitude and the frequency of the free frequency fit is closer to the remnant
overtone at later times. We speculate that the reason this does not happen for
the remnant perturbation case is that the overtone amplitude is smaller, and
becomes unresolved at earlier times. 
We also recall that, in the instantaneously changing background
case shown above (Fig.~\ref{fig:mode_id_example_FFT}) there is no evidence for any nonmode content,
and the overtone amplitude is extracted cleanly.
We further discuss these different
initial data and background models in Appendix \ref{sec:Physical_case}. 

\begin{figure}[h]
\centering
\includegraphics[width=0.5\textwidth]{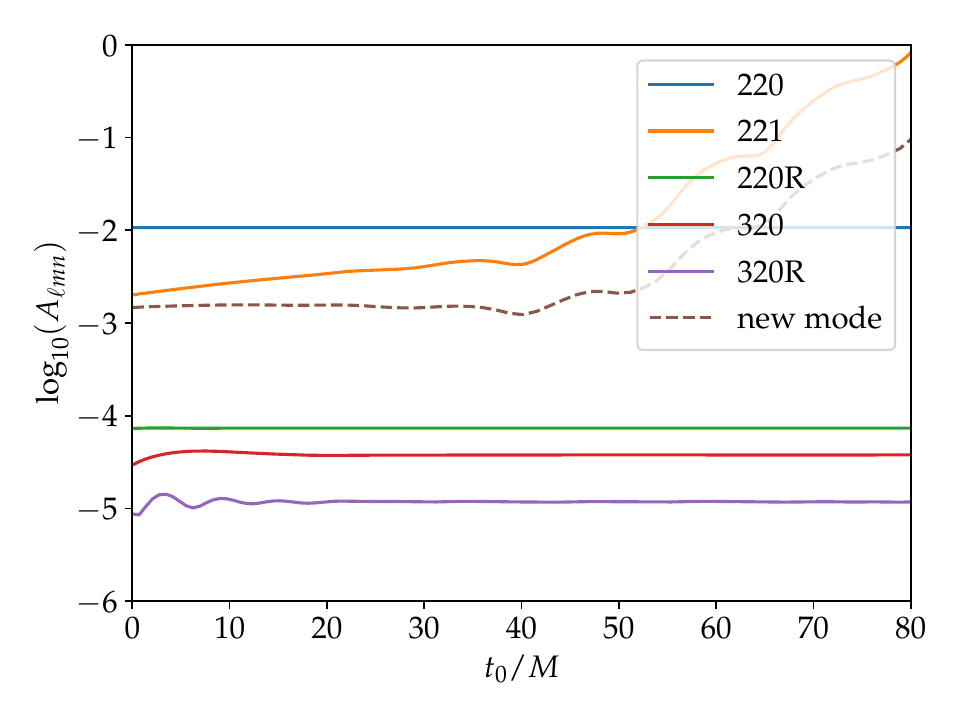}
    \caption{Fitted quasinormal mode amplitudes relative to $t=0$ for a remnant black hole with spin $a/M=0.7$ and a background changing such that $\Delta M/M = 1\%$. These amplitudes are defined in Eq.s~\eqref{eq:amp_ref_time} and ~\eqref{eq:fixed-freq-model}. We apply filters and do independent fits for fundamental modes [i.e. with $N=1$ in Eq.~\eqref{eq:fixed-freq-model}]. We fit the first overtone together with the $n=2$ and $3$ overtones (setting $N = 3$). 
We also show a fit with a quasinormal mode with a frequency of $\omega = 0.526 + 0.221i$ (labeled new mode) that comes from a free frequency fit, as opposed to being a quasinormal mode frequency of the remnant black hole.
    }
\label{fig:amps_with_fit_time}
\end{figure}

As anticipated above, when fitting explicitly for the $(221)$ overtone we find a time dependence in the measured amplitude not present
in the other quasinormal modes. This is shown in Fig.~\ref{fig:amps_with_fit_time}, where we plot the nominally time-independent amplitudes with fit time (defined relative to an arbitrary reference time by Eq.~\eqref{eq:amp_ref_time}) extracted using the fixed-frequency model in Eq.~\eqref{eq:fixed-freq-model}. At early times the inferred amplitude of the $(221)$ can be seen to increase by a factor of a few at roughly an exponential rate. (At late times, it increases even more rapidly, but this is behavior is expected since this mode decays much faster than the other modes, and eventually becomes buried in the noise.) This means that this residual mode can be interpreted as a $(221)$ mode with a shifted decay rate and indeed, picking a new quasinormal mode frequency from a free frequency fit at $t=10M$ as our template, we find that the extracted amplitude becomes more flat in fit time (labeled by ``new mode'' in Fig.~\ref{fig:amps_with_fit_time}). 
We find that the shift in decay rate is spin dependent, increasing for an increasing rate of background change.

Although the measurement is somewhat ambiguous, we choose to extract an
overtone amplitude at the time when the amplitude is smoothest with variations
in the fitting window (and the free frequency most closely matches the
overtone) to obtain to the results described in Sec.~\ref{sec:Results}, 
with the caveat that they
do not represent a fully resolved overtone. For the
example in Fig.s~\ref{fig:free_freq_overtonea} and ~\ref{fig:amps_with_fit_time}, this corresponds to $t = (25-30)M$.  We find
that, when considering higher black hole spins, the region of maximally flat
amplitude in fit time moves to later times. This makes sense, since the
timescale of the background change is half the timescale of the perturbing
mode, which increases with increasing spin. In general, we find that the most
flat region for the overtone fit corresponds approximately to the time $t = 2 /
\mathfrak{I}\omega_{220}$. 

\subsection{Changing frequency model}
\label{sec:changing_freq}

With the goal of understanding the nonmode content to the ringdown signal indicated by the above results, we now consider a changing frequency model, along the lines of the one accounting for the third-order changing background effect \cite{Redondo-Yuste:2023ipg}. 
For simplicity, we will focus on a example with zero angular momentum, in particular, considering a nonspinning black hole with an $\ell =2, \ m=0, \ n=0$ quasinormal perturbation. This means that the instantaneous frequency will only evolve according to the change in the black hole's mass. 
Modifying the changing frequency model of Ref.~\cite{Redondo-Yuste:2023ipg} to integrate the phase up to each time instead of using the instantaneous frequency, we have
\begin{align}
\label{eq:integrated_freq}
\phi(t) &= \int_{t_0}^t \omega_{200}(t') dt' = \int_{t_0}^t \frac{m_1\omega_{200}}{m(t')} dt'\\
&= \left(\frac{\omega_{200} m_1}{-2 \mathfrak{I}\omega}\right) \log \left( \frac{  \exp(-2\mathfrak{I}\omega t)-\Delta M}{ \exp(-2\mathfrak{I}\omega t_0)-\Delta M} \right), \nonumber \\
\Psi &= \Tilde{A}\left[1+\Tilde{Q}\frac{\delta m(t)}{m_2-m_1}\right]\exp\left(i\phi(t) \right)
.
\end{align}
We introduce the start time $t_0$ as a fitting parameter, since we do not know the time of propagation from the horizon to null infinity with high accuracy. Here $\omega_{200}$ is the fundamental mode of the initial black hole with mass $m_1$, and $m(t) = m_1 + \delta m(t) = m_1 + \Delta m (1-\exp(2 \mathfrak{I}\omega_{200}t))$. The other free parameters are $\Tilde{A}\in \Bbb{C}$ and $\Tilde{Q}\in \Bbb{C}$ \footnote{In \cite{Redondo-Yuste:2023ipg}, the value of $\Tilde{Q}$ is fit to data, and should be the same for all cases. Since there is some uncertainty in this fit, we make $\Tilde{Q}\in \Bbb{C}$ a free parameter to give this model the best chance of matching the data. In fact, we find that setting $\Tilde{Q}=0$ in our integrated phase model does not significantly affect the fit performance (while it does seem to make an improvement in the adiabatic approximation model from Ref.~\cite{Redondo-Yuste:2023ipg}).}. 

\begin{figure}[h]
\centering
\includegraphics[width=0.5\textwidth]{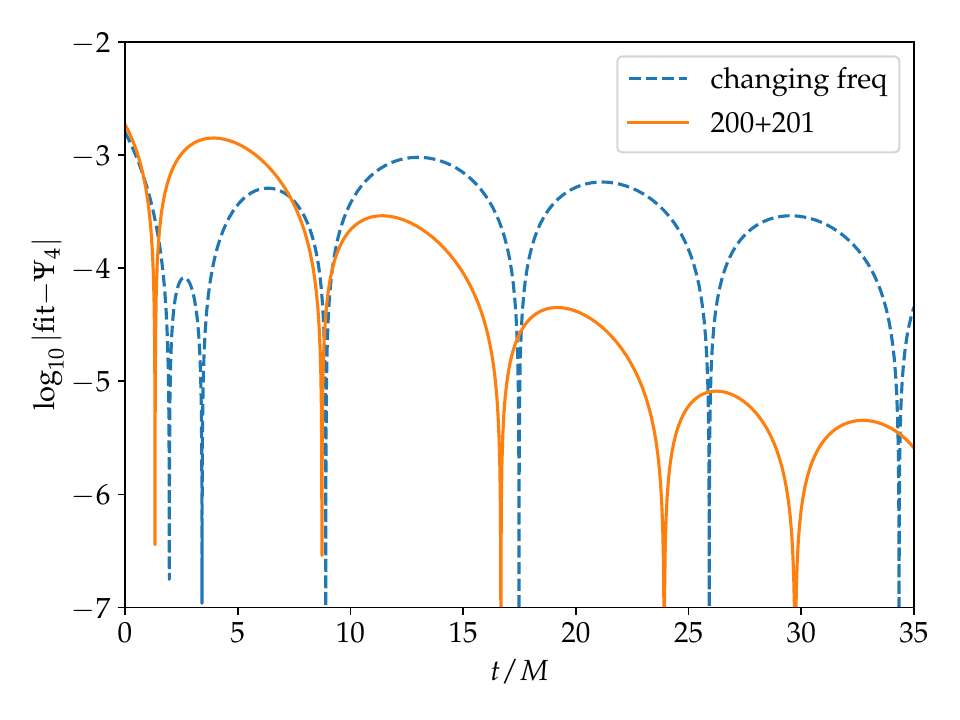}
    \caption{A comparison of a changing frequency model [Eq.~\eqref{eq:integrated_freq}] to a fixed-frequency
    model with fundamental and overtone $(200)+(201)$.
    We show the real part of the residual of two fits to $\Psi_4$. As only the relative values of the two residuals
    are important, the overall scale of the $y$ axis is arbitrary.
    }
\label{fig:residual_changing_freq_fit}
\end{figure}

We consider an example on a evolving background with $\Delta M /M \approx 2\%$ and 
find the same behavior as in the spinning cases in the overtone amplitude (as was shown in Fig.~\ref{fig:amps_with_fit_time})
as well as when performing a free frequency fit after filtering out the fundamental and retrograde modes (as shown in Fig.~\ref{fig:filtered_allmodes_changingbkgd}).
This example is thus representative of the behavior described above. We compare how well the changing frequency model
 does in fitting the ringdown in this example compared to a fixed-frequency model with two modes: $(200)$ and $(201)$.
As shown in Fig.~\ref{fig:residual_changing_freq_fit}, we find the changing frequency model captures the gravitational wave signal better during the first $10M$, but does worse at later times when we expect the signal to asymptote to a sum of quasinormal modes (while
the changing frequency model will asymptote to only the fundamental mode). 

\begin{figure}%
    \centering
    \includegraphics[width=0.5\textwidth]{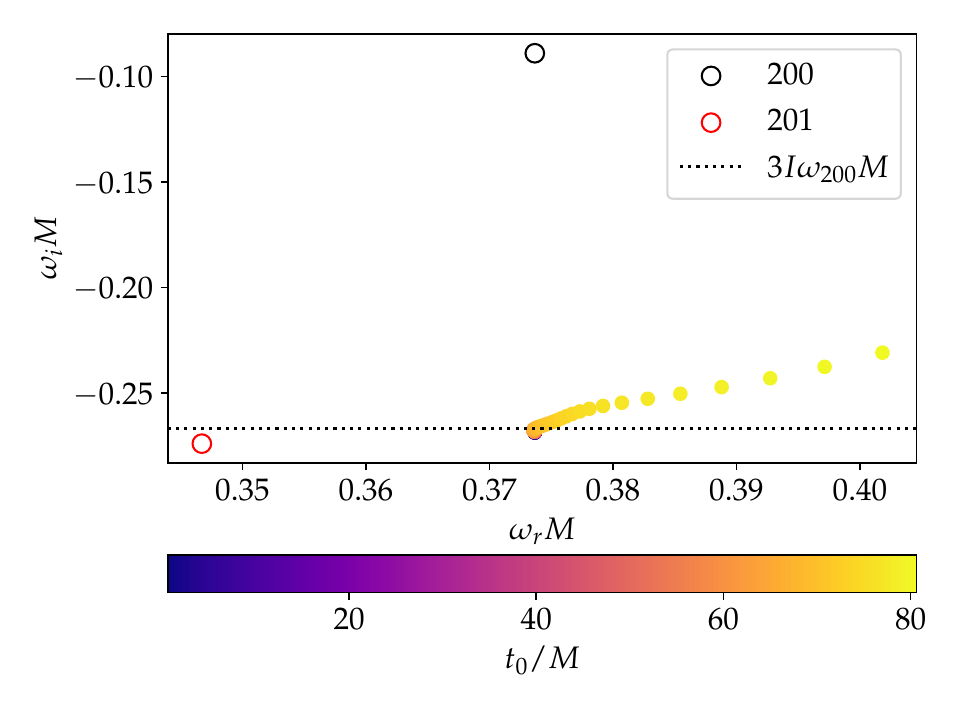} 
    \caption{
    The results of a free frequency fit on the changing frequency model given by Eq.~\eqref{eq:integrated_freq} after filtering out the fundamental mode.
    The points with $t<50M$ points are stacked closely together and have $\omega_I\approx \mathfrak{I}\omega_{201} \approx 3 \mathfrak{I}\omega_{200}$.
    }
    \label{fig:changing_freq_fit_free_freq}%
\end{figure}

Finally, we can illustrate how a changing frequency component would affect our fixed-frequency mode analysis by applying it directly to the  waveform from Eq.~\eqref{eq:integrated_freq}. Using the best-fit parameters from the above example and first applying a $(200)$ quasinormal mode filters, and then fitting a free frequency, we obtain the results shown in Fig.~\ref{fig:changing_freq_fit_free_freq}.
For the first $\approx 60M$ (after which the results are not well resolved), the analysis identifies a mode with real frequency $\approx \mathfrak{R}\omega_{200}$, 
but imaginary frequency $\approx \mathfrak{I}\omega_{201}$. However, we note that, in fact, $\mathfrak{I}\omega_{201} \approx 3 \mathfrak{I}\omega_{200}$ (to within $2.5\%$). Some component with the latter decay rate is expected in Eq.~\eqref{eq:integrated_freq} for a fundamental quasinormal mode with frequency $\omega_{200}$ that is interacting with (multiplying in the evolution equation) a changing mass/spin component to the black hole background that goes like $\sim \exp(2\mathfrak{I}\omega_{200}t)$. 

Based on these results, we can conclude that in the ringdown on the evolving background, for both the spherically symmetric and spinning cases,
there are components that one can attribute both to the presence of a changing fundamental frequency, as well to the excitation of an overtone. Because of the similar, and short, decay times of these two features, they are hard to disentangle.
One the one hand, one has a decay rate of $ 3\mathfrak{I}\omega$ for the cubic effect of a mode interacting with the changing black hole properties at twice the mode's decay rate.
On the other hand, in the WKB approximation \cite{Yang:2012he}, the imaginary component of the quasinormal frequency changes with overtone number as $\propto (n+1/2)$, which 
means that, in general, $\mathfrak{I}\omega_{n=1} \approx 3 \mathfrak{I}\omega_{n=0}$; this holds to $<1\%$ for the $a/M=0.7$ case shown in Fig.~\ref{fig:filtered_allmodes_changingbkgd}, which shows similar behavior at early times.

\section{Discussion and Conclusion}
\label{sec:discussion}

\subsection{Summary of main results}
\label{sec:summary_results}

In this paper, we have studied the nonlinear effect on a ringdown gravitational
wave signal due to the changing black hole mass and spin attendant in the
ringdown process. Our analysis is based on promoting the mass and spin in the
Teukolsky equation governing linear metric perturbations to be functions of
time, as prescribed by the flux of gravitational wave energy and angular
momentum into the black hole due to a quasinormal mode. Though somewhat
simplistic, it allows us to isolate and quantify one particular nonlinear
effect in black hole ringdown, among many that may be present, and test the
consistency of assuming linear evolution in a given regime.  We find that,
beginning from single least-damped quasinormal mode, this effect causes a phase
shift and amplitude diminishment in the fundamental mode, while exciting
overtone, retrograde, and higher $\ell$ modes. In addition, we find evidence
that as the black hole background evolves at earlier times (e.g. $t\lesssim 30M$ for $a/M = 0.7$), 
there is a component to the signal
that cannot be described purely as a superposition of remnant quasinormal modes, 
likely due to a changing frequency contribution.

\subsection{Comparison to binary black hole ringdown}
\label{sec:lessons_for_analysis}

The results described here can also be taken as a self-consistency check for when a black hole ringdown
signal is in the linear regime: a ringdown signal can only be said to be described purely
by a sum of quasinormal modes only to the degree to which the nonlinear effects described here (as well 
as others) can be ignored.
We can compare the magnitude of the effects found here to the ringdown signals
found in comparable mass binary black hole mergers. For a range of binaries,
the relative excitation of the first overtone and retrograde mode have been
measured by Ref.~\cite{Cheung:2023vki}. We present those amplitudes, together with
our estimate of the contribution due to AIME for a case with a remnant spin of
$a/M=0.7$ (computed as strain amplitudes). 
For the latter, we include the values both when using initial data corresponding
to a fundamental quasinormal of the remnant black hole and corresponding to a 
quasinormal mode of the initial black hole. 
\begin{widetext}
\begin{center}
\begin{table}[ht]
\caption{Estimate of AIME in comparison with total excitation from numerical relativity (NR). The errors shown are due to an uncertainty in reference time.}
\label{tab:comparing_exc_phys}
\begin{tabular}{ c|c|c } 
 & $h_{221}/h_{220}$ & $h_{220R}/h_{220}$ \\
\hline
NR & $4 \pm 1$ & $10^{-4} \pm 10^{-4} $\\ 
AIME (remnant ID) & $(0.34 \pm 50\%)\times \frac{\Delta M/M}{0.01} $ & $(2.1\times 10^{-2} \pm 5\%)\times \frac{\Delta M/M}{0.01}$ \\ 
AIME (initial BH ID) & $(0.4 \pm 50\%)\times \frac{\Delta M/M}{0.01} $ & $(3.5\times 10^{-3} \pm 5\%)\times \frac{\Delta M/M}{0.01}$ \\ 
\end{tabular}
\end{table}
\end{center}
\end{widetext}

In Table~\ref{tab:comparing_exc_phys}, $\Delta M/M$ is the mass change in the linear regime due to the absorption of the fundamental $(220)$ mode. 
As described above, estimating the
magnitude of the least-damped quasinormal mode from the postpeak gravitational
wave signal implies a $\sim1 \%$ in the black hole mass and $\sim 6\%$
change in the spin, though there is some ambiguity in connecting the time at
null infinity with that at the black hole horizon. This is a factor of few
smaller than the change of mass of the final apparent horizon found in binary
black hole simulations.

In this example, we estimate that the AIME effect excites the first overtone
amplitude at $(5-20)\%$ the value found in binary black hole mergers. On the other hand, we
estimate the retrograde excitation due to this effect to be $(20-75)\times$
the value fit to numerical relativity simulations. Of course, in a binary
black hole merger, various quasinormal modes---in addition to the least-damped one---will be significantly excited by the dynamics
of the merger, which will both add (constructively or destructively) with those
generated by nonlinear effects in the ringdown, as well as affect how the black hole
mass and spin changes at early times. 
We note, for example, that the ratio between the change in the apparent horizon
mass and spin from the simulation shown Fig.~\ref{fig:mass_spin_change_in_t}
is noticeably different from that expected from a $(220)$ mode.
It would be interesting for future work to calculate the horizon flux due to 
a superposition of quasinormal modes and see how the resulting evolution 
of black hole mass and spin affect the results found here. 

We can also compare the significance of this third-order effect with the second-order mode doubling effect. The strain amplitudes of mode doubled modes are fit
in Refs.~\cite{Zhu:2024rej,Mitman:2022qdl,Redondo-Yuste:2023seq,Cheung:2023vki} for
various numerical relativity simulations. Here we compare to
Ref.~\cite{Cheung:2023vki}, where they find that these modes are of order
$h_{220\times220}/h_{220}^2 = (0.125-0.175)$, and
$h_{220\times320}/(h_{220}h_{320}) = (0.4-0.6)$. The mode doubled amplitudes
are also calculated analytically in \cite{Ma:2024qcv}, where for $a/M = 0.7$,
they find $h_{220\times220}/h_{220}^2 = (0.1233)$. 
As an example, in the binary black hole merger described in Sec.~\ref{sec:mass_spin_change_due_to_qnm} (with mass ratio of $1.2$, remnant spin $a/M = 0.69$ and low eccentricity \cite{GieslerEvent}), 
the measured mode amplitudes are $h_{220} \approx 1$ and $h_{320} \approx 0.03$
\cite{Cheung:2022rbm}.
To compare these effects, we can use ratios of the time-integrated strain squared, 
which would roughly correspond to the square of the signal-to-noise ratio (SNR) assuming white noise.
That is, the ratio of the SNR of a particular mode $(\ell m n)$ to that of another mode $(\ell'm'n')$ is given by
\begin{align}
    \frac{\text{SNR}_{\ell m n}}{\text{SNR}_{\ell'm'n'}} &= \sqrt{\frac{\int_{0}^{\infty}dt \left|h_{\ell m n}\right|^2}{\int_{0}^{\infty}dt \left|h_{\ell' m' n'}\right|^2}} = \frac{|A_{\ell m n}|\sqrt{-\mathfrak{I}\omega_{\ell' m' n'}}}{|A_{\ell' m' n'}|\sqrt{-\mathfrak{I}\omega_{\ell m n}}},
\end{align}
where $A_{\ell m n}$ are the amplitudes at $t=0$. Taking the signal from the
peak, with amplitudes from Table~\ref{tab:comparing_exc_phys} and the example
case referenced above, this gives
$\text{SNR}_{221}^{\text{AIME}}/\text{SNR}_{220\times220} \approx (2-3) $ and
$\text{SNR}_{220_R}^{\text{AIME}}/\text{SNR}_{220\times320} \approx  (0.3-2)$.
Recall that in the overtone case we do not extract a stable amplitude until
late times. If we calculate instead the SNR for the overall change in signal
with a single quasinormal mode, we find
$\text{SNR}^{\text{AIME}}/\text{SNR}_{220\times220} \approx 0.2- 0.5$.
According to this na\"ive approximation, the third-order AIME effect 
is comparable in magnitude to second-order mode doubling.

Following Ref.~\cite{Purrer:2019jcp}, we calculate the maximum ringdown SNR such that the pure $220$ mode signal and the signal with our AIME result included are indistinguishable, or $\rho$ where the mismatch $\mathcal{M}$ satisfies
\begin{align}
    \label{eq:max_snr}
    \mathcal{M}(h_{220}, h_{\text{AIME}}) < \frac{D}{2\rho^2}
    ,
\end{align}
where $D$ is the number of model parameters. We calculate mismatch in the time domain using
\begin{align}
\mathcal{M}(f(t),g(t)) = 1-\frac{\int^{t_f}_{t_0}(f(t)^*g(t)) dt}{\sqrt{\int_{t_0}^{t_f}(f(t)^*f(t))dt \int_{t_0}^{t_f}(g(t)^*g(t)) dt}}
.    
\end{align}
We approximate $\mathcal{M}(h_{220}, h_{\text{AIME}})$ by the mismatch between the signals with and without a changing background, where $t_0$ is set to the lookback time and $t_f=60M$. We find the range of SNRs for indistinguishability, 
\begin{equation}
    \rho < 60 \left(\sqrt{\frac{D}{4}}\right) \left(\frac{0.01}{\Delta M/M}\right)
.
\end{equation}
For reference, the postpeak SNR of GW150914 is $\sim 14$ \cite{Isi:2019aib}, which would correspond to $\Delta M/ M \approx 4\%$ in the above.
According to our leading-order approximation, this effect is only starting to be relevant for current ringdown analyses from the peak.

\subsection{Limitations and directions for future work}
\label{sec:limitations}

In this study, we have focused on a particular nonlinear effect in black hole
ringdown, due to the changing black hole mass and spin, while ignoring other
nonlinear effects. Our numerical evolution code simulated a changing
background by changing the evolution operators everywhere on the hyperbolic
time slice together, which introduces introduces an unphysical
instantaneousness, and ignores the mass and angular momentum of the
gravitational perturbation outside the black hole horizon.  As well, we have
only calculated the leading-order contribution to the black hole mass and spin
change, and ignored how backreaction may affect this.  While doing so allowed
us to isolate, and estimate in a straightforward manner, the effect of the
changing black hole mass and spin, it would be desirable for future work to
perform a fully relativistic calculation of black hole ringdown that would
quantify this effect (as well as other nonlinear ringdown effects). This could
be done using numerical relativity simulations of isolated black holes
with a perturbation whose amplitude is varied~\cite{East:2013mfa,Sberna:2021eui,Zhu:2024rej}, but in the spirit of the
calculation performed here, beginning with initial data where the perturbation closely
approximates a single quasinormal mode of a Kerr black hole. 

In the setup studied here, we found evidence for nonmode content in the
ringdown, which we attempted to account for using a simple changing frequency
model. A corresponding effect has not so far been identified in the ringdown of
binary black hole simulations and may be artificially magnified by our setup,
or may be hard to disentangle from the nonlinearities in the merger.  However,
it might be beneficial, for future work, to further develop such models so that
they can better capture mildly nonlinear effects in black hole ringdown
signals.

\section{Acknowledgments}

T.M. would like to thank Gregorio Carullo, Luis Lehner, Eric Poisson, and Laura Sberna for useful discussion and particularly Mark Ho Cheung for the suggestion that a changing fundamental frequency can result in a frequency close to the first overtone.
S. M. would like to thank Matthew Giesler, Hengrui Zhu, Keefe Mitman, and Saul Teukolsky for useful discussion.
W.E., T.M., and S.M. acknowledge support from a Natural Sciences and Engineering Research
Council of Canada Discovery Grant and an Ontario Ministry of Colleges and
Universities Early Researcher Award. This research was supported in part by
Perimeter Institute for Theoretical Physics. Research at Perimeter Institute is
supported in part by the Government of Canada through the Department of
Innovation, Science, and Economic Development and by the Province of Ontario
through the Ministry of Colleges and Universities. Calculations were performed
on the Symmetry cluster at Perimeter Institute.

\newpage
\onecolumngrid
\appendix
\section{Change in a black hole mass and spin due to the absorption of gravitational waves}
\label{App:analytic_calc_dA}

Here we review how to compute the change in the mass and spin of a black hole from the absorption of a mode solution to the Teukolsky equation.
This problem was first examined in \cite{Hawking:1972hy} and elaborated upon in \cite{Teukolsky:1974yv,1986PhRvD..33..915P,1986bhmp.book.....T}. 
A review and extension of those earlier results to the time domain was performed in \cite{Poisson:2004cw}. 
Despite these earlier works, for completeness and to set our notation, we provide a detailed review of the relevant calculations here.

First we compute how the black hole area changes due to the absorption of linearized gravitational waves within the NP formalism.
We then use the first law of black hole thermodynamics, along with the Fourier decomposition of the linearized gravitational perturbations, to relate the change in the black hole area to the change in the mass and spin of the hole.

\subsection{Transforming to the Hawking-Hartle tetrad\label{sec:coordinates}}

We begin by deriving the Hawking-Hartle tetrad \cite{Hawking:1972hy}; see also Sec. 98(d) of Ref.~\cite{Chandrasekhar:1985kt}.
Following Hawking, Hartle, and Chandrasekhar,
we make use of the Newman-Penrose formulation of the 
Einstein equations \cite{Newman:1961qr}.
The metric for a Kerr black hole in ingoing Kerr-Schild coordinates is
\begin{align}
   \label{eq:ingoing-metric}
   g_{ab}dx^adx^b 
   &=
   \left(1-\frac{2Mr}{\Sigma_{\rm BL}}\right)dv^2
   -
   2dvdr
   +
   \frac{4aMr\sin^2\vartheta}{\Sigma_{\rm BL}}dvd\varphi
   +
   2a\sin^2\vartheta drd\varphi
   \nonumber\\
   &
   -
   \Sigma_{\rm BL}d\vartheta^2
   -
   \left(
      a^2+r^2+2Mr\frac{a^2}{\Sigma_{\rm BL}}\sin^2\vartheta
   \right)
   \sin^2\vartheta
   d\varphi^2
   ,
\end{align}
where $\Sigma_{\rm BL} \equiv r^2 + a^2\cos^2\vartheta$, $\Delta_{\rm BL} \equiv \left(r-r_+\right)\left(r-r_-\right)$, and $r_{\pm}\equiv M \pm \sqrt{M^2 - a^2}$ are the inner and outer horizon radii.
The metric \eqref{eq:ingoing-metric} follows from the Boyer-Lindquist metric via the coordinate transformation
\begin{subequations}
\label{eq:to-df-coords}
\begin{align}
    dv &=
    dt 
    +
    \frac{r^2+a^2}{\Delta_{\rm BL}} dr
    ,\\
    d\vartheta
    &=
    d\phi
    +
    \frac{a}{\Delta_{\rm BL}}dr
    .
\end{align}
\end{subequations}
The inverse metric in the transformed coordinates is
\begin{align}
   g^{ab}\partial_a\partial_b
   &=
   -
   \frac{a^2\sin^2\vartheta}{\Sigma_{\rm BL}}\partial_v^2
   -
   \frac{2\left(r^2+a^2\right)}{\Sigma_{\rm BL}}\partial_v\partial_r
   -
   \frac{\Delta_{\rm BL}}{\Sigma_{\rm BL}}\partial_r^2
   -
   \frac{2a}{\Sigma_{\rm BL}}\partial_v\partial_{\varphi}
   -
   \frac{2a}{\Sigma_{\rm BL}}\partial_v\partial_{\vartheta}
   -
   \frac{1}{\Sigma_{\rm BL}}\partial_{\vartheta}^2
   -
   \frac{\csc^2\vartheta}{\Sigma_{\rm BL}}\partial_{\varphi}^2
   .
\end{align}
We see that on the black hole horizon, $r$ is a null coordinate.
If $a=0$, $v$ is also a null coordinate, but when $a\neq0$ it is timelike.
The Kinnersley tetrad \cite{Kinnersley:1969zza} in ingoing Kerr-Schild coordinates is 
\begin{subequations}
\begin{align}
   l^{a}_{(K)}
   \partial_{a}
   &=
   \frac{2\left(r^2+a^2\right)}{\Delta_{\rm BL}}\partial_v
   +
   \partial_r
   +
   \frac{2a}{\Delta_{\rm BL}}\partial_{\varphi}
   ,\\
   n^{a}_{(K)}\partial_{a}
   &=
   -
   \frac{\Delta_{\rm BL}}{2\Sigma_{\rm BL}}\partial_r
   ,\\
   m^{a}_{(K)}\partial_{a}
   &=
   \frac{1}{2^{1/2}\left(r+ia\cos\vartheta\right)}
   \left(
      ia\sin\vartheta\partial_v
      +
      \partial_{\vartheta}
      +
      \frac{i}{\sin\vartheta}\partial_{\varphi}
   \right)
   .
\end{align}
\end{subequations}
This tetrad is not regular on the black hole horizon.
We transform to the Hawking-Hartle tetrad \cite{Hawking:1972hy}, which is regular, via the rotation
\begin{align}
    \label{eq:K_to_HH_tetrad}
   l_{(HH)}^a
   &\equiv
   \frac{\Delta_{\rm BL}}{2\left(r^2+a^2\right)}
   l_{(K)}^a
   ,\\
   n_{(HH)}^a
   &\equiv
   \frac{2\left(r^2+a^2\right)}{\Delta_{\rm BL}}
   n_{(K)}^a
   .
\end{align}
We then have
\begin{subequations}
\begin{align}
   l_{(HH)}^a\partial_a
   &=
   \partial_v
   +
   \frac{\Delta_{\rm BL}}{2\left(r^2+a^2\right)}
   \partial_r
   +
   \frac{a}{r^2+a^2}\partial_{\varphi}
   ,\\
   n_{(HH)}^a\partial_a
   &=
   -
   \frac{r^2+a^2}{\Sigma_{\rm BL}}\partial_r
   ,\\
   m_{(HH)}^a\partial_a
   &=
   \frac{1}{2^{1/2}\left(r+ia\cos\vartheta\right)}
   \left(
      ia\sin\vartheta\partial_v
      +
      \partial_{\vartheta}
      +
      \frac{i}{\sin\vartheta}\partial_{\varphi}
   \right)
   .
\end{align}
\end{subequations}

\subsection{Properties of the ingoing Kerr-Schild metric and the Hawking-Hartle tetrad\label{sec:hh_properties}}
The Hawking-Hartle tetrad is well defined on the horizon, and we will use it to calculate horizon fluxes.
The Ricci-rotation coefficients and Weyl scalars are listed in \cite{Ripley:2020xby}.
As $\kappa=0$, $l_{(HH)}^a$ is pregeodesic (see \cite{Chandrasekhar:1985kt} Sec. 9);
that is, $l^a\nabla_al^b\propto l^b$. 
On the black hole horizon, $l_{(HH)}^a$ is tangent to the null
generators of the black hole horizon; that is, it can
be written as \cite{Hawking:1972hy,Bardeen:1973gs}
\begin{align}
   l_{(HH)}^a
   =
   \frac{dx^a}{dv}
   .
\end{align}
This follows from Eq.~\eqref{eq:to-df-coords} and the fact that $\Delta_{\rm BL}=0$ on the black hole horizon
\begin{align}
   \frac{dx^a}{dv}
   \partial_a
   &=
   \frac{d v}{d v}\partial_v
   +
   \frac{d r}{d v}\partial_r
   +
   \frac{d\vartheta}{d v}\partial_{\vartheta}
   +
   \frac{d\varphi}{d v}\partial_{\varphi}
   \nonumber\\
   &=
   \partial_v
   +
   \frac{a}{r^2+a^2}\partial_{\varphi}
   .
\end{align} 
Thus on the black hole horizon the NP derivative operator 
$D_{(HH)}\equiv l_{(HH)}^a\partial_a$ is
\begin{align}
   \label{eq:hh_tetrad_D_der_horizon}
      D_{(HH)}
      =
      \frac{d}{dv}
      .
\end{align}
Moreover, it is straightforward to see that 
\begin{align}
   \xi_{(v)}^a\partial_a
   \equiv
   \partial_v
   ,\qquad
   \xi_{(\varphi)}^a\partial_a
   \equiv
   \partial_{\varphi}
\end{align}
are Killing vectors for the spacetime. 
On the black hole horizon we can then write
\begin{align}
   \label{eq:hh_tetrad_l_horizon}
      l_{(HH)}^a
      =
      \xi_{(v)}^a
      +
      \Omega_H\xi_{(\varphi)}^a
      ,
\end{align}
where the angular velocity of the horizon $\Omega_H$ is $\Omega_H\equiv\frac{a}{r_+^2+a^2}=\frac{a}{2Mr_+}$.
We note that $l^a_{(HH)}$ and $m^a_{(HH)}$ commute with each other on the inner and outer horizons, 
\begin{align}
    l^a_{(HH)}\nabla_am_{(HH)}^b
    =
    m^a_{(HH)}\nabla_al_{(HH)}^b
    .
\end{align}

\subsection{Change in the black hole area}

The area of the black hole horizon is
\begin{align}
   \label{eq:black-hole-area}
   A(v)
   =
   \int d\varphi d\theta \sqrt{s}
   ,
\end{align}
where $s_{AB}$ is the induced metric on the black hole horizon.
We can write
\begin{align}
   s_{AB}
   \equiv
   g_{ab}e^a_Ae^b_B
   ,
\end{align}
where $e^a_A$ are the (spacelike, essentially angular) tangent vectors on the horizon, 
and $d\varphi d\theta $ are the angular coordinates on the horizon.
In ingoing Kerr-Schild coordinates, we have
\begin{align}
   e_{(1)}^a
   \partial_a
   =
   \frac{dx^a}{d\vartheta}
   \partial_a
   =
   \partial_{\vartheta}
   ,
   \qquad
   e_{(2)}^a
   \partial_a
   =
   \frac{dx^a}{d\varphi}
   \partial_a
   =
   \partial_{\varphi}
   .
\end{align}
This implies that 
\begin{align}
   s_{AB}d\theta^Ad\theta^B
   =
   g_{\vartheta\vartheta}d\vartheta^2
   +
   g_{\varphi\varphi}d\varphi^2
   ,\qquad
   \sqrt{s}
   =
   \left(g_{\vartheta\vartheta}g_{\varphi\varphi}\right)^{1/2}
   .
\end{align}
   The $e_A^a$ can be converted to the NP vectors $m^a$ by first
defining a complex transformation vector $e_A$ to obtain
a complex vector $e^a$ that satisfies
\begin{align}
   e^a
   =
   e^Ae^a_A
   ,\qquad
   g_{ab}e^ae^b
   =
   0
   ,\qquad
   g_{ab}e^a\bar{e}^b
   =
   -1
   .
\end{align}
From the form of the Hawking-Hartle tetrad, we see that on the black hole horizon ($\Delta_{\rm BL}=0$) we can choose a complex $e_A$ and $A$ such that
\begin{align}
   e^a
   =
   m_{(HH)}^a
   +
   Al_{(HH)}^a
   .
\end{align}
We note that $l^a_{(HH)}$ commutes with the angular vectors $e_A^a$
\begin{align}
   l_{(HH)}^a\nabla_ae^b_A
   =
   e^a_A\nabla_al_{(HH)}^b
   .
\end{align}
Using all these relations, we have [dropping the $(HH)$ suffix to make the equations less cluttered]
\begin{align}
   \frac{dA}{dv}
   &=
   \int d\varphi d\theta  \sqrt{s}\left(\frac{1}{\sqrt{s}}\frac{d\sqrt{s}}{dv}\right)
   \nonumber\\
   &=
   \int d\varphi d\theta  \sqrt{s} \left(\frac{1}{2}s^{AB}l^c_{(HH)}\nabla_cs_{AB}\right)
   \nonumber\\
   &=
   -
   \int d\varphi d\theta  \sqrt{s} 
   g_{ab}\left(
      \left(
         \bar{m}^c
         m^b
         +
         m^c
         \bar{m}^b
      \right)
      \left(\nabla_cl^a\right)
   \right)
   \nonumber\\
   &=
   -
   \int d\varphi d\theta  \sqrt{s} \left(\rho + \bar{\rho}\right)
   .
\end{align}
We have used that $l^c\nabla_cl^a\propto l^a$, $l^am_a=0$, $l^al_a=0$, 
and the definition for $\rho\equiv\bar{m}^a\left(\nabla_al^b\right) m_b$
(see Ref.~\cite{Chandrasekhar:1985kt} \S8).
In summary, we have that the change the black hole area is the surface
integral over the horizon of twice the real part of the
Newman-Penrose scalar $\rho$ in the Hawking-Hartle tetrad
in ingoing Kerr-Schild coordinates,
\begin{align}
   \label{eq:hh_tetrad_change_bh_area_general}
  \frac{dA}{dv}
  =
  -
  \int d\varphi d\theta  \sqrt{s}\left(\rho_{(HH)}+\bar{\rho}_{(HH)}\right)
  .
\end{align}
We note that the real part of $\rho$ corresponds to the expansion of the null congruence $l^a$ \cite{Chandrasekhar:1985kt}, which describes the change rate of the congruence's cross-sectional area \cite{Poisson:2009pwt}.

\subsection{The Newman-Penrose equations for the change in the black
hole area\label{sec:change_are_np}}

Note that Eq.~\eqref{eq:hh_tetrad_change_bh_area_general} is technically nonperturbative---we do not rely on expanding the solution to a fixed order in perturbation theory.
To relate $dA/dv$ to solutions to the Teukolsky equation ($\Psi_0^{(1)}$, we perturbatively solve the NP equations; see Refs.~\cite{Hawking:1972hy} or \cite{Chandrasekhar:1985kt} Sec. 98(d).

We start with Ref.~\cite{Chandrasekhar:1985kt} (310a) and (310b), 
\begin{align}
   D\rho
   &=
   \left(\epsilon + \bar{\epsilon}\right)\rho
   +
   \rho^2 + \sigma\bar{\sigma}
   +
   \left(\bar{\delta} - 3\alpha - \bar{\beta} + \varpi\right)\kappa
   -
   \tau\bar{\kappa}
   +
   \Phi_{00}
   ,\\
   D\sigma 
   &=
   \left(\rho+\bar{\rho}+3\epsilon-\bar{\epsilon}\right)\sigma
   +
   \left(\delta - \bar{\alpha} - 3\beta - \tau + \bar{\varpi}\right)\kappa
   +
   \Psi_0
   .
\end{align}
Here $\Phi_{00}\equiv - (1/2)R_{ab}l^al^b = - (1/2)T_{ab}l^al^b$ represents
the matter field contribution; $\Phi_{00}^{(0)}=0$ since
the background is vacuum Kerr.
We will work with a Hawking-Hartle tetrad to all orders in perturbation theory.
That is, we assume $l^a$ is pregeodesic 
($\kappa=0$), and we assume $D=d/dv$ to all orders in perturbation theory.
The NP equations then are
\begin{align}
   \frac{d\rho_{(HH)}}{dv}
   &=
   \left(\epsilon_{(HH)} + \bar{\epsilon}_{(HH)}\right)\rho_{(HH)} 
   +
   \rho_{(HH)}^2 + \sigma_{(HH)}\bar{\sigma}_{(HH)} 
   +
   \Phi_{00}
   ,\\
   \frac{d\sigma_{(HH)}}{dv}
   &=
   \left(
      \rho_{(HH)}+\bar{\rho}_{(HH)}+3\epsilon_{(HH)}-\bar{\epsilon}_{(HH)}
   \right)\sigma_{(HH)}
   +
   \Psi_{0(HH)}
   .
\end{align}
As $\sigma_{(HH)}^{(0)}=0$ and $\rho_{(HH)}^{(0)}=0$ on the black
hole horizon, and $\epsilon_{(HH)}^{(0)}$ is real, 
to linear order in perturbation theory we see that 
\begin{align}
   \frac{d\rho_{(HH)}^{(1)}}{dv}
   &=
   2\epsilon_{(HH)}^{(0)}\rho_{(HH)}^{(1)}
   ,\\
   \frac{d\sigma_{(HH)}^{(1)}}{dv}
   &=
   2\epsilon_{(HH)}^{(0)}
   \sigma_{(HH)}^{(1)}
   +
   \Psi_{0(HH)}^{(1)}
   .
\end{align}
Note that we have assumed that $\Phi_{00}^{(1)}=0$.
This is a reasonable assumption (at least to linear order in perturbation theory) as the matter stress-energy tensor
is typically quadratic in the matter fields so the linear perturbation would be
zero.
The only solution to the first equation is $\rho_{(HH)}^{(1)}=0$
(with the boundary condition $\rho_{(HH)}^{(1)}=0$ at $v=\infty$), 
so we expand the $\rho$ equation to second order to obtain
\begin{align}
   \frac{d\rho_{(HH)}^{(2)}}{dv}
   &=
   2\epsilon_{(HH)}^{(0)}\rho_{(HH)}^{(2)}
   +
   \left|\sigma_{(HH)}^{(1)}\right|^2
   +
   \Phi^{(2)}_{00(HH)}
   .
\end{align}
Setting $\rho_{(HH)}^{(2)}=0$ and $\sigma_{(HH)}^{(1)}=0$ at $v=\infty$, we have
\begin{subequations}
\begin{align}
   \label{eq:eq_for_rho2_hh_tetrad}
  \rho_{(HH)}^{(2)}(v,\vartheta,\varphi)
  &=
  -
  \int_v^{\infty}dv'\mathrm{exp}\left[
        2\epsilon_{(HH)}^{(0)}\left(v-v'\right)
     \right]
  \left(
     \left|\sigma_{(HH)}^{(1)}\left(v',\vartheta,\varphi\right)\right|^2
     +
     \Phi^{(2)}_{00(HH)}\left(v',\vartheta,\varphi\right)
  \right)
  . \\
   \label{eq:eq_for_sigma1_hh_tetrad}
  \sigma_{(HH)}^{(1)}\left(v,\vartheta,\varphi\right)
  &=
  -
  \int_v^{\infty}dv'\mathrm{exp}\left[
     2\epsilon_{(HH)}^{(0)}\left(v-v'\right)
  \right]
   \Psi_{0(HH)}^{(1)}\left(v',\vartheta,\varphi\right)
    .
\end{align}
\end{subequations}
We see that from $\Psi_{0(HH)}^{(1)}$ we can obtain $\sigma_{(HH)}^{(1)}$ and then $\rho_{(HH)}^{(2)}$.
Finally, we see that $\rho_{(HH)}^{(0)}=\rho_{(HH)}^{(1)}=0$ on the black hole horizon,
and $\rho_{(HH)}^{(2)}$ is real.
Then from Eq.~\eqref{eq:hh_tetrad_change_bh_area_general}, we have to leading order that \cite{Hawking:1972hy}
\begin{align}
      \frac{dA^{(2)}}{dv}
      =
      -
      2\int d\varphi d\theta  \sqrt{s} \rho^{(2)}\left(v,\vartheta,\varphi\right)
      .
\end{align}
\subsection{Relating change in the black hole area to the 
change in the black hole mass and spin\label{sec:change_mass_spin}}

The black hole area $A=8\pi M r_+$ is given by Eq.~\eqref{eq:black-hole-area}.

Note that $\sqrt{s}$ on the black hole horizon is the same in ingoing Kerr-Schild (as we are using here) and Eddington-Finkelstein coordinates (which are used in the evolution code referenced in the main text). Defining $J\equiv a M$, we then have 
\begin{align}
   dA
   &=
    \frac{8 \pi  \left(-JdJ+2 M^2 r_+dM\right)}{\sqrt{M^4-J^2} }
    .
\end{align}
From this, we conclude that
\begin{align}
   \frac{d^2A}{dvd\Omega}
   =
   \frac{8 \pi}{\sqrt{M^4-J^2} }\left(2 M^2 r_+ \frac{d^2M}{dvd\Omega}  -J \frac{d^2J}{dvd\Omega}\right)
   .
\end{align}
We consider perturbations in frequency space, that is, perturbations
of the form
\begin{align}
   \Psi_0^{(1)}\left(v,r,\vartheta,\varphi\right)
   =
   \int d\omega \sum_m
   e^{im\varphi-i\omega v}
   \tilde{\Psi}_0\left(\omega,r,\vartheta,m\right)
   .
\end{align}
As $\xi_{(v)}^a$ and $\xi_{(\varphi)}^a$ are Killing vectors,  we have \cite{Sberna:2021eui,Teukolsky:1974yv} 
\begin{align}
   \label{eq:energy_angular_momentum_relation_fs}
      dJ
      =
      m\frac{\mathfrak{R}\omega}{\left|\omega\right|^2}dE
\end{align}
for perturbations $dJ$ and $dE$ of the spacetime angular momentum
and spacetime energy. 
With Eq.~\eqref{eq:energy_angular_momentum_relation_fs}
and $E=M$, $J=a M$, we have
\begin{align}
   \frac{d^2A}{dvd\Omega} &=
   \frac{8 \pi  \left(2 M r_+ | \omega | ^2 -a m \mathfrak{R}\omega 
   \right)}{\sqrt{M^2-a^2} | \omega | ^2}
   \frac{d^2M}{dvd\Omega}
    .
\end{align}
Note that the sign changes for certain values of $\omega$. The sign of the mass change depends on the sign of \\$\left[ \frac{|\omega|^2}{\mathfrak{R}\omega} - m  \Omega_H \right]$. This is the superradiant condition in the complex case.
Inverting this equation, we obtain Eq.~\eqref{eq:mass_change_from_area_change} in the main text.
Using Eq.~\eqref{eq:energy_angular_momentum_relation_fs} and Eq.~\eqref{eq:mass_change_from_area_change}, we have 
\begin{align}
    \label{eq:angular_momentum_change_from_area_change}
      \frac{d^2J}{dvd\Omega}
      &= 
      \frac{ \sqrt{M^2-a^2} m \mathfrak{R}\omega }{8 \pi  \left(2 M r_+ | \omega | ^2 -a m \mathfrak{R}\omega 
   \right)}
      \frac{d^2A}{dvd\Omega}
      .
\end{align}
Using $dJ = Mda + a dM$ and Eq.s ~\eqref{eq:mass_change_from_area_change} and ~\eqref{eq:angular_momentum_change_from_area_change}, we obtain Eq.~\eqref{eq:a_change_from_area_change} in the main text.

\subsection{Change in the black hole mass and spin for a gravitational 
quasinormal mode \label{sec:change_mass_spin_qnm}}

We consider a quasinormal mode solution to $\Psi_0$,
\begin{align}
   \Psi_0\left(v,r,\vartheta,\varphi\right)
   =
   \mathcal{A}
   \times
   e^{im\varphi - i\omega v}
   R\left(r\right)S\left(\vartheta\right)
   ,
\end{align}
where $\mathcal{A}$ is the amplitude of the quasinormal mode,
and $S$ is a spin-weighted spheroidal harmonic \cite{Teukolsky:1973ha}.
We normalize the solution so that $R(r_+)=1$.
On the black hole horizon, we have
\begin{align}
    \label{eq:epsilon_def}
   \epsilon^{(0)}_{(HH)}
   &=
   \frac{
      \left(M^2-a^2\right)^{1/2}
   }{
      4Mr_+
   }
   .
\end{align}
We can now compute $d^2A/(dvd\Omega)$ for this solution.

First we compute $\sigma^{(1)}_{(HH)}$ using 
Eq.~\eqref{eq:eq_for_sigma1_hh_tetrad},
\begin{align}
   \sigma^{(1)}_{(HH)}\left(v,\vartheta,\varphi\right)
   &=
   -
   \mathcal{A}
   S\left(\vartheta\right)
   e^{im\varphi}
   \int_v^{\infty}
   dv'
   \mathrm{exp}\left[2\epsilon_{(HH)}^{(0)}\left(v-v'\right)-i\omega v'\right]
   \nonumber\\
   &=
   \mathcal{A}
   \frac{
      S\left(\vartheta\right)
      e^{im\varphi-i\omega v}
   }{
      2\epsilon^{(0)}_{(HH)}+i\omega 
   }
   .
\end{align}
From Eq.~\eqref{eq:eq_for_rho2_hh_tetrad} we have
\begin{align}
   \rho^{(2)}_{(HH)}\left(v,\vartheta,\varphi\right)
   =
   -
   \left|\mathcal{A}\right|^2
      \left|S\left(\vartheta\right)\right|^2
   \int_v^{\infty}
   dv'
   \frac{
      \mathrm{exp}\left[
         2\epsilon_{(HH)}^{(0)}\left(v-v'\right)
         +
         2\left(\mathfrak{I}\omega\right)v'
      \right]
   }{
        \left|2 \epsilon^{(0)}_{(HH)} -\mathfrak{I}\omega  \right|^2
      +
      \left|\mathfrak{R}\omega \right|^2
   }
   .
\end{align}

From Eq.~\eqref{eq:eq_for_A2_hh_tetrad}, we conclude that
\begin{align}
   \frac{d^2A^{(2)}}{dv}
   =
   2
   \left|\mathcal{A}\right|^2
   \int d\varphi d\theta\sqrt{s}
      \left|S\left(\vartheta\right)\right|^2
   \int_v^{\infty}
   dv'
   \frac{
      \mathrm{exp}\left[
         2\epsilon_{(HH)}^{(0)}\left(v-v'\right)
         +
         2\left(\mathfrak{I}\omega\right)v'
      \right]
   }{
      \left|2 \epsilon^{(0)}_{(HH)} -\mathfrak{I}\omega  \right|^2
      +
      \left|\mathfrak{R}\omega \right|^2
   }
   .
\end{align}
In our case (and in the evolution code coordinates), we have an induced horizon metric $s_{AB}$ such that: 
\begin{align}
    \label{eq:our-metric}
    \sqrt{s} 
    &= 2M r_+\sin (\theta ) .
\end{align}

Using the fact that the spin-weighted spheroidal harmonics are normalized
to have unit norm over the $\vartheta$ integral \cite{Teukolsky:1973ha} [Eq. (2.8)],
we can integrate over the sphere to obtain 
\begin{align}
   \label{eq:change_in_area_for_qnm}
      \frac{dA^{(2)}}{dv}
      =
      4\pi
      \left|\mathcal{A}\right|^2 
      \int_v^{\infty}
      dv' 2 M r_+ 
      \frac{
         \mathrm{exp}\left[
            2\epsilon_{(HH)}^{(0)}\left(v-v'\right)
            +
            2\left(\mathfrak{I}\omega\right)v'
         \right]
      }{
         \left|2 \epsilon^{(0)}_{(HH)} - \mathfrak{I}\omega  \right|^2
      +
      \left|\mathfrak{R}\omega \right|^2
      } .
\end{align}

Now we take $M \neq M(v)$ and $a\neq a(v)$, making this a leading level calculation only. Carrying out this integral assuming a fixed background gives Eq.~\eqref{eq:area-change-psi0-mode-sol} in the main text.
Setting $\mathfrak{I}\omega = 0$, we find this matches the pure mode result in \cite{Poisson:2004cw} [Eqs. (5.14) and (5.22)] up to a coordinate choice.

Say we know the black hole mass at time $v=v_1$.
We can then compute the black hole area at time $v=v_2$ 
using Eq.~\eqref{eq:area-change-psi0-mode-sol} 
\begin{align}
     \triangle A
      &\equiv
      A_2
      -
      A_1 
       = 2\pi
      \left|\mathcal{A}\right|^2
       M r_+
      \frac{\mathrm{exp}\left[2 \mathfrak{I}\omega v_2\right]-\mathrm{exp}\left[2 \mathfrak{I}\omega v_1\right]}
      { \mathfrak{I}\omega \left(\epsilon^{(0)}_{(HH)} -\mathfrak{I}\omega\right) \left((2 \epsilon^{(0)}_{(HH)} - \mathfrak{I}\omega
   )^2+\mathfrak{R}\omega^2\right)}
   .
\end{align}
The approximate change in the black hole mass and spin over that
interval are, from Eqs.~\eqref{eq:mass_change_from_area_change} and \eqref{eq:a_change_from_area_change}: 
\begin{align}
   \label{eq:change_in_M_for_qnm_triangle}
   \triangle M
   \approx&
   \frac{\sqrt{M_0^2-a_0^2} | \omega | ^2}{8 \pi  \left(2 M_0 r_+ | \omega | ^2 -a_0 m \mathfrak{R}\omega 
   \right)}
   \triangle A
   ,\\
   \label{eq:change_in_a_for_qnm_triangle}
   \triangle a
   \approx&
   \frac{\sqrt{M_0^2-a_0^2}  \left(a_0 | \omega | ^2+m \mathfrak{R}\omega\right)}{8 \pi M_0 \left(2 M_0 r_+ | \omega | ^2 -a_0 m \mathfrak{R}\omega 
   \right)}
   \triangle A
   .
\end{align}

From these formulas it may at first appear that in the extremal limit 
$dM/dv = da/dv = 0$ for this approximation. However, looking more closely at the expression for $dA/dv$, we see that, in the real frequency case, this factor cancels out (since $\epsilon^{(0)}_{(HH)} \propto \sqrt{M^2-a^2}$). In the extremal case, quasinormal modes approach real frequencies. Note that, apart from the extremal case or some fine-tuning of the perturbation, we have $\Delta a = 0$ only for $a_0 = 0$ and a perturbation with $m=0$.


\section{Physical perturbing amplitude}
\label{App:motivation_perturbing_amp}

In order to apply our flux formulas, we need to relate a strain amplitude at null infinity to the amplitude of $\Psi_0$ on the black hole horizon. Here we derive a relation between these. First, we relate the strain amplitude to that of the Weyl scalar $\Psi_4$ at infinity. Then we use numerical solutions to the Teukolsky equation to relate the amplitude at infinity to one on the horizon in the Kinnersley tetrad. We then use a result from \cite{Berens:2024czo} to relate $\Psi_4$ to $\Psi_0$ on the horizon. We then rotate to the well-defined Hawking-Hartle tetrad. Finally, we account for the effect of choosing a particular time slicing.

\subsection{Strain amplitudes and Weyl scalars}

Since here we are concerned with the amplitude of the Weyl scalars, we estimate the amplitude of $\Psi_4$. From the relation $\Psi_4^K(r\rightarrow\infty) = 1/2(d^2h/dt^2) $, for a single $(220)$ quasinormal mode we have 
\begin{align}
\label{eq:strain_to_psi4}
    |\Psi_4^K (r\rightarrow\infty)| = \frac{ |h_{220}\omega_{220}^2|}{2}\frac{1}{r}
    ,
\end{align}
where we recall that the definition of $h_{\ell m n}$ is given by Eq.~\eqref{eq:hlmn}.

\subsection{Moving from null infinity to the horizon}

In order to solve the Teukolsky equation, the ``long-range potential'' in the Teukolsky equation can be removed (see \cite{Ripley:2022ypi} Eq.~6), and a scaling with $\rho^{(0)} = -1/(r-ia\cos\theta)$ introduced (\cite{Teukolsky:1973ha} Table 1---this is necessary to have a separable Teukolsky equation). This gives new variables $\psi$. These $\psi$ are well defined on both the horizon and null infinity. Explicitly,
\begin{align}
\label{eq:code_convention_psi}
    \Psi^{K}_4 &\equiv \frac{1}{r}\Delta_{\rm BL}^{-s} \rho^{(0)4} \psi_4\\
    &^{r\gg M}\approx r^{-1}\psi_4\\
    \Psi^{K}_0 &\equiv \frac{1}{r}\Delta_{\rm BL}^{-s} \psi\\
    &^{r\gg M}\approx r^{-5}\psi_0
    .
\end{align}
We relate the value of $\psi_0$ at null infinity to its value at the horizon. We do this by numerically calculating the radial eigenfunctions of the Teukolsky equation in our evolution hyperbolic coordinates using \cite{Ripley:2022ypi} [this code uses the Kinnersley tetrad, with the variables described in Eq.~\eqref{eq:code_convention_psi}]. We define this ratio
\begin{align}
    \label{eq:to_horizon_psi0}
    C_{\text{to-horizon}} = \left|\frac{\psi_4(r_+)}{\psi_4(\infty)}\right|
    .
\end{align}

\subsection{Relating Weyl scalars $\Psi_0$ and $\Psi_4$ }

We use a result from Ref.~\cite{Berens:2024czo} using metric reconstruction to relate $\Psi_0$ and $\Psi_4$ at infinity 
in Boyer-Lindquist coordinates and the Kinnersley tetrad. According to this calculation, a solution [Eq. (2.43) in Ref.~\cite{Berens:2024czo}]
\begin{align}
    \Psi_4^{K} = \rho^{(0)4}e^{-i\omega t +im\phi} R^{(-2)} \ _2S_{\ell m}
\end{align}
is equivalent to a solution [Eq. (2.44) in Ref.~\cite{Berens:2024czo}]
\begin{align}
    \Psi_0^{K} = \frac{4 D'}{\mathfrak{C}'}e^{-i\omega t +im\phi} R^{(2)}\ _{2}S_{\ell m} + \frac{48i\omega^*M}{\mathfrak{C}'^*}e^{+i \omega^* t -im\phi} R^{(2)}\ _{2}S_{\ell -m}^*
    ,
\end{align}
where $D'$ and $\mathfrak{C}'$ are versions of the Starobinsky constant, see Ref.~\cite{Berens:2024czo} for the complete definitions. Here $^*$ indicates the complex conjugate. The prograde $m$ mode is the first term and the retrograde $-m$ mode is the second term. So we have two modes corotating with the black hole with the same frequency, described by two different quasinormal modes. From Ref.~\cite{Berens:2024czo}, these radial functions are normalized such that (for ingoing physical solutions)
\begin{align}
    \left|R^{s}(r\rightarrow r_+)\right|\rightarrow \Delta_{\rm BL}^{-s}
    .
\end{align}
Taking the norm of the corresponding solutions, we have
\begin{align}
\label{eq:TS_identity_Kinnersley2}
    |\Psi_4^{K}(\infty)| &= \mathcal{A}\left|\rho^{(0)4}\Delta_{\rm BL}^2\right|\\
    |\Psi_0^{K}(\infty)| &= \sqrt{\int d\Omega \Psi_0^{K} \Psi_0^{K*} }\\
    &= \mathcal{A}\Delta_{\rm BL}^{-2}\sqrt{\left|\frac{4D'}{\mathfrak{C}'}\right|^2 +\left|\frac{48 \omega M}{\mathfrak{C}'}\right|^2}
    .
\end{align}
Note that the cross term (corresponding to the second-order mode) disappears in the norm and so does not change the black hole area. 
Thus we get the relation between the norms [in terms of the regularised variables from Eq.~\eqref{eq:code_convention_psi}]
\begin{align}
    |\psi_0(r_+)| = |\psi_4(r_+)|\sqrt{\left|\frac{4D'}{\mathfrak{C}'}\right|^2 +\left|\frac{48 \omega M}{\mathfrak{C}'}\right|^2}
    .
\end{align}
Then we have
\begin{align}
    \left|\Psi_0^{K}(r_+)\right| &= 
    \label{eq:inf_to_hor_K}
    \left|\left(\frac{1}{r}\Delta_{\rm BL}^{-2}\right)_{r_+}\sqrt{\left|\frac{4D'}{\mathfrak{C}'}\right|^2 +\left|\frac{48 \omega M}{\mathfrak{C}'}\right|^2} C_{\text{to-horizon}} \left(r\Delta_{\rm BL}^{-2} \rho^{(0)-4}\right)_{\infty} \Psi_4^{K}(\infty)\right|
    .
\end{align}
In our case, $C_{\text{to-horizon}}$ is the same for both the $m$ prograde mode and the $-m$ retrograde mode (both corotating with frequency $\omega$).
Recall that this relation applies only to the Kinnersley tetrad. 

\subsection{Relating to Hawking-Hartle tetrad}

The two tetrads are related through a transformation known as a tetrad rotation; specifically, in the language of Chandrasekhar \cite{Chandrasekhar:1985kt}, they are related by a type III rotation. From Ref.~\cite{Chandrasekhar:1985kt} Sec. 347, a type III rotation sends the tetrad components $l \rightarrow A^{-1} l $, $n \rightarrow A n$, $m \rightarrow e^{i\theta} m $, $\Bar{m} \rightarrow e^{-i\theta} \Bar{m}$. The Weyl scalars change according to
\begin{align}
        \label{eq:tetrad_rotation_psi4}
    &\Psi_0 \rightarrow A^{-2} e^{2i\theta} \Psi_0,
    &\Psi_4 \rightarrow A^2 e^{-2i\theta} \Psi_4 ,
\end{align}
where $A$ and $\theta$ are defined by the specific rotation. The translation from the Hawking-Hartle to the Kinnersley tetrad corresponds to $\theta_{HH\rightarrow K} = 0$ and $A_{HH\rightarrow K} = \frac{\Delta_{\rm BL}}{2(r^2+a^2)}$ from \eqref{eq:K_to_HH_tetrad}. 
Using the argument from Ref.~\cite{Teukolsky:1974yv} [Eq. (4.43)], if Weyl scalars in different tetrads are equal on the horizon, then it is correct to use the same flux through the horizon, and so we can use a tetrad rotation to calculate the input for our flux formulae. 
From Eq. \ref{eq:tetrad_rotation_psi4}, $\Psi_0$ in the Hawking-Hartle tetrad is related to the Kinnersley tetrad by
\begin{align}
    \Psi_0^{K}(r_+) &= \left(
    A_{HH\rightarrow K}^{-2}  e^{2i \theta_{HH\rightarrow K}} \Psi_0^{HH} (r_+)\right) \\
    &= \left(\frac{4(r^2+a^2)^2}{\Delta_{\rm BL}^2}\right)_{r_+}\Psi_0^{HH} (r_+)
    .
\end{align}
And finally, using Eqs.~\eqref{eq:inf_to_hor_K} and \eqref{eq:strain_to_psi4},
\begin{align}
\label{eq:phys_amp}
    \left|\Psi_0^{(HH)}(r_+)\right| 
    &= \left|\frac{C_{\text{to-horizon}}}{16 M^2r_+^3} \sqrt{\left|\frac{4D'}{\mathfrak{C}'}\right|^2 +\left|\frac{48 \omega M}{\mathfrak{C}'}\right|^2} \left(\frac{ |h_{220}\omega^2|}{2}\right)\right|
    .   
\end{align}

\subsection{Lookback time}

Equation~\eqref{eq:phys_amp} relates the amplitude at the horizon in the Hawking-Hartle tetrad to the amplitude of the strain at infinity
at a given time, for our particular choice of time slicing, and assuming a quasinormal mode that has existed for all time.
However, physically we are interested in the scenario where we have some transient perturbation in the vicinity of the black hole
(roughly, say, in the vicinity of the light ring) and we wish to synchronize a gravitational wave signal at infinity with 
a flux at the horizon arising from the same perturbation. This introduces a ``lookback time'' ambiguity.

The quantity given in Eq.~\eqref{eq:phys_amp} is the amplitude at the horizon in the Hawking-Hartle tetrad when the signal at infinity is assumed to be composed of a quasinormal mode. However, assuming the whole spacetime is described by a quasinormal mode, we expect the signal at the horizon to be composed of quasinormal modes at an earlier time. At some coordinate time the behavior in the bulk is assumed to be linear. This behavior (we approximate the peak at the light ring) sends signatures to null infinity and to the black hole horizon. The linear behavior should be apparent first at the horizon (in our coordinate choice). Since we take this behavior to be linear, we can account for this effect by considering an increase in initial amplitude corresponding to the change in start time for the linear behavior. 

As a rough estimate of the lookback time and its uncertainty, we can take a
null geodesic starting at the light ring and calculate the coordinate time of
propagation to null infinity $T_{\infty}$ and to the horizon $T_{r_+}$. These
propagation times are calculated according to
Appendix~\ref{App:propagation_times} in the same coordinates
as those for the calculation of
$C_{\text{to-horizon}}$. The difference in those times corresponds to an estimate of the
lookback time, the coordinate time difference to synchronize
a signal at infinity with the flux at the horizon.
The change in the amplitude of a quasinormal mode associated with this is a factor of
\begin{align}
\label{eq:look-back-time}
    C_{\text{lookback}} = e^{\mathfrak{I}\omega \left(T_{\infty}-T_{r_+}\right)}.
\end{align}
Taking an uncertainty in the radial peak of the quasinormal mode of $1M$ (in the Boyer-Lindquist radius) corresponds to an uncertainty in the amplitude of a factor of 2.

\section{Propagation times}
\label{App:propagation_times}
In order to estimate the lookback time connecting a gravitational perturbation near the black hole to a time at null infinity in 
our particular evolution coordinates, 
we calculate the propagation times from the light ring to the horizon and null infinity.
The evolution coordinates $\{t,R,\theta,\varphi\}$ are related to the usual ingoing Eddington-Finkelstein coordinates
by $R=1/r$ and $t = v-r-4M\log(r)$ (for more discussion, see, for example, Appendix C of \cite{Ripley:2020xby}).

For the case of $a/M = 0.7$, if we consider a radially outward, equatorial, null geodesic originating near the light ring at $r=r_{\rm LR}\ ^{+M}_{-M/5}$, we find the coordinate time to reach null infinity is $T_{\infty}=(9 \pm 5) M$.
We get similar results if we numerically evolve some for narrowly peaked pulse in $\Psi_4$ at the same position and find the time it takes to reach null infinity.

\begin{figure}[h]
\centering
\includegraphics[width=0.5\textwidth]{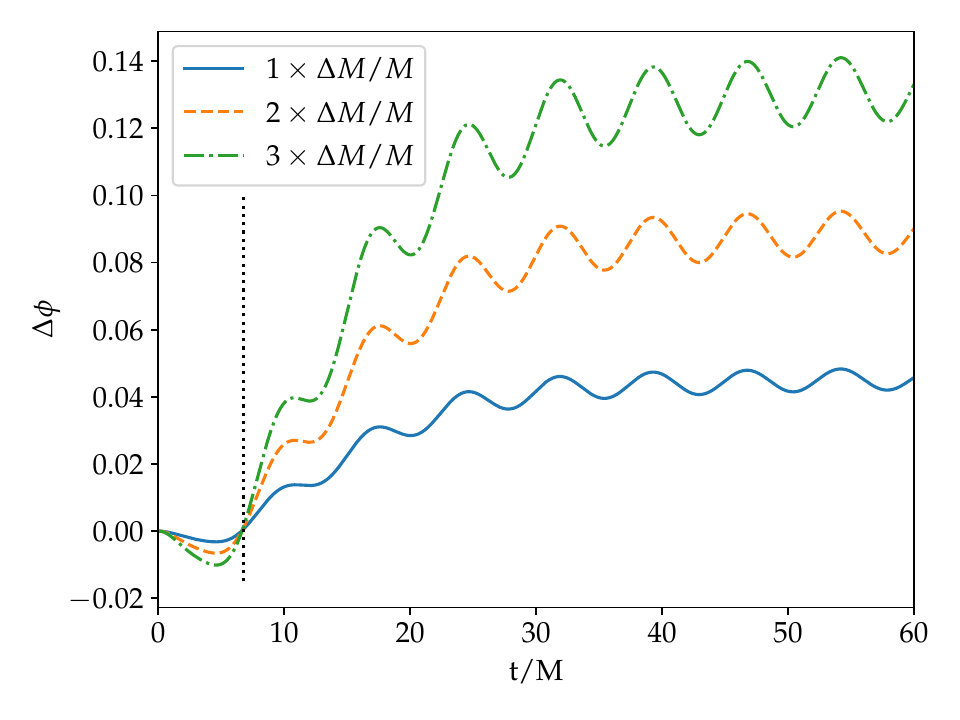}
\caption{The change in signal at infinity is shown for $a/M = 0.7$ and various degrees of changing black hole mass, with the lowest mass change $\Delta M/M = 0.7\%$. $\Delta\phi(t)$ is the phase difference in $\Psi_4$ between the changing background case and the fixed remnant background case as a function of time. As can be seen from the figure, this asymptotes to a constant value, as described in Sec.~\ref{sec:Results}. The black dotted line is located at $t=6.75M$ and shows where the different cases start to diverge.}
\label{fig:phase_change_diff_dMs}
\end{figure}

We can compare this estimate of the lookback time to the calculation using a quasinormal mode. The total phase change of the signal (relative to the original quasinormal mode) at infinity due to a changing background is calculated for several values of $\Delta M$ in Fig.~\ref{fig:phase_change_diff_dMs}. The phase difference begins to grow after $t\approx7M$, which also provides an estimate
of how long it takes for the part of the quasinormal mode in the near zone that is sensitive to the black hole dynamics
to propagate to null infinity. This is comparable, though slightly smaller than the estimate based on null geodesics. 

\section{Comparison of different quasinormal initial data}
\label{sec:Physical_case}

In Secs.~\ref{sec:Results} and \ref{sec:ResultsII}, we consider a setup where the initial data corresponded exactly to a quasinormal mode of the final black hole, with the initial black hole changing mass and spin due to the absorption of this perturbation to reach these final parameters. In some sense this can be considered a consistency check---if a quasinormal mode of the remnant is present at some time, then we should see this third-order effect. For comparison, here we consider initial data corresponding to a quasinormal mode the initial black hole.
Considering these types of initial data also allows us to compare the excitation of different modes when the black hole evolves according to Eqs.~\eqref{eq:adiabatically_changing_background_M} and \eqref{eq:adiabatically_changing_background_a} versus instantaneously changing black hole background (i.e. fixing the mass and spin to the remnant values). 

\begin{figure}%
    \centering
    \subfloat[\centering Instantaneously changing background]{{\includegraphics[width=8cm]{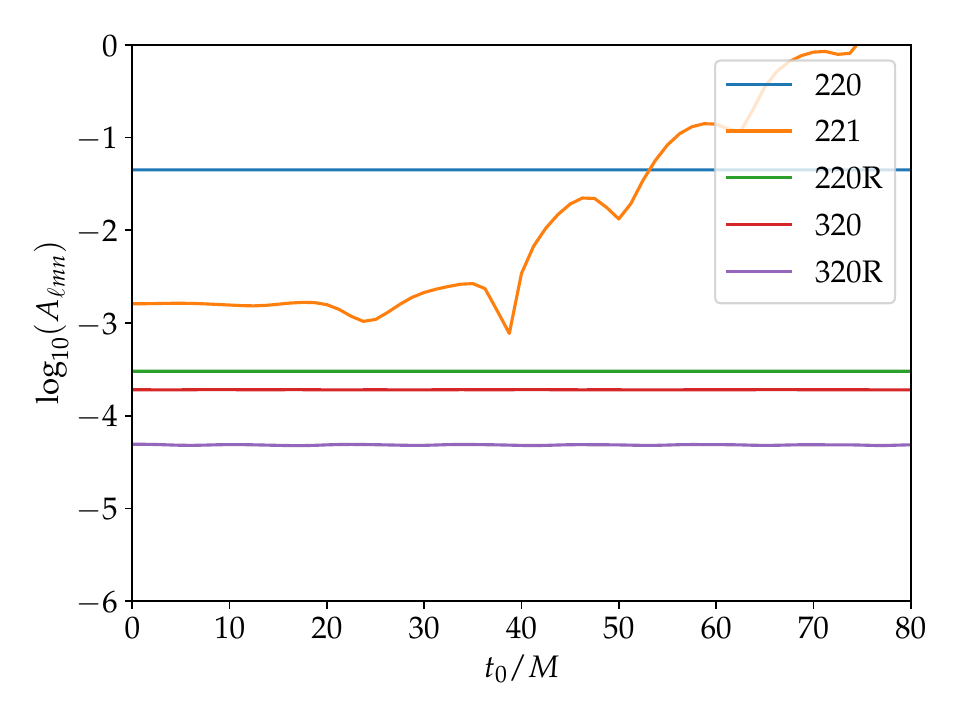} }}%
    \qquad
    \subfloat[\centering Evolving background]{{\includegraphics[width=8cm]{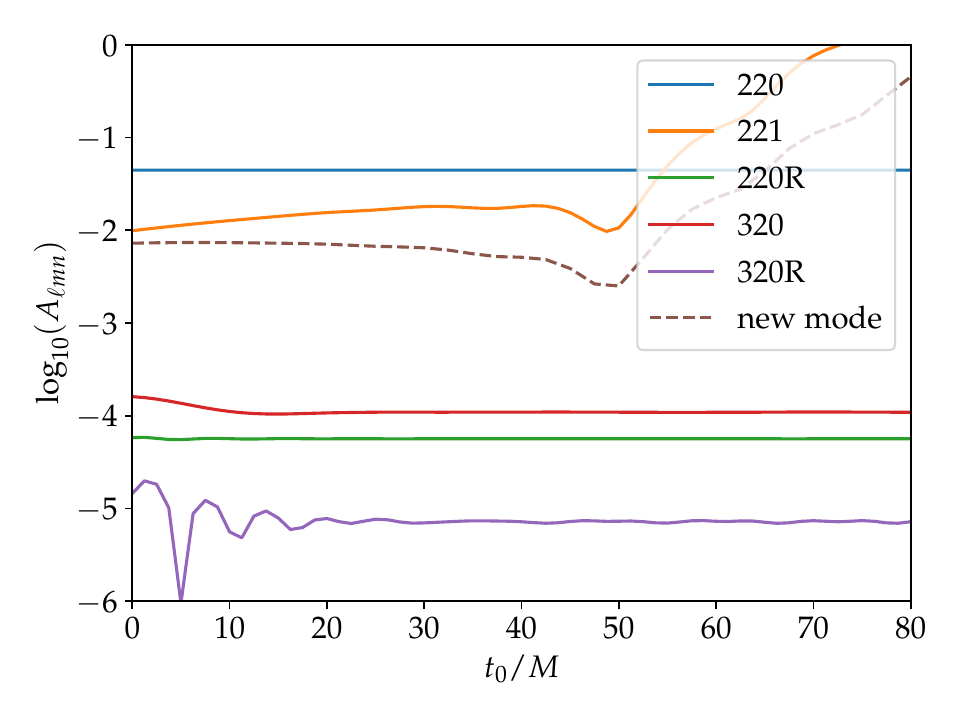} }}%
    \caption{Comparing linear evolution of the same initial perturbation on two
	different backgrounds. 
        In both cases the initial perturbation is a
        quasinormal mode of a black hole with $M_{\text{BH}} = (1-0.01) M$ and spin
        $a_{\text{BH}} = 0.7(1-0.063)M$.
        (b) The case when the black hole evolves from these parameters 
        to have mass $M$ and spin $a/M=0.7$ [according to Eqs.\eqref{eq:adiabatically_changing_background_M}
        and \eqref{eq:adiabatically_changing_background_a}], while (a) shows the case where
        the black hole has the latter parameters for all time.
	We see that the changing background enhances overtones, while the
	retrograde modes are relatively suppressed. We include a fit for a mode
	frequency  $\omega = (\mathfrak{R}\omega_{220}) + i
	(\mathfrak{I}\omega_{221}+0.0236)$ for the evolving background case. The new
	mode has constant amplitude at $t\sim (0-20)M$, but the physical $(221)$ mode
	has a more constant amplitude at $t_0\sim (30- 45)M$. 
    }%
    \label{fig:ComparingBackgrounds}%
\end{figure}

In Fig.~\ref{fig:ComparingBackgrounds}, we compare the amplitudes of the modes
extracted in these two cases (cf. Fig.~\ref{fig:amps_with_fit_time}). We find
that using an evolving background enhances overtone excitation, while
suppressing retrograde excitations. We also find that using these initial data
allows us to more confidently extract an overtone compared to using a
quasinormal mode of the final black hole.  Here it seems that this new
mode (with frequency determined by a free frequency fit at $t=10M$, as
described in Sec.~\ref{sec:ResultsII}) has a more constant amplitude, and so
fits the data better, at early times. However, in this case, the physical
$(221)$ mode has approximately constant amplitude in the $t_0\sim (30- 45)M$
region. This region corresponds to the region where a free frequency fit
returns something close to the first overtone frequency.  Recall that
Fig.~\ref{fig:free_freq_overtone} shows free frequency fits after filtering out
fundamental modes for both initial data cases and an evolving background. When
using a quasinormal of the initial black hole we more confidently identify an
overtone amplitude for longer and find a frequency closer to the remnant
overtone at later times, compared to using a quasinormal mode of the final
black hole.  We speculate that this is simply due to the fact that the
overtone excitation is larger and hence can be resolved longer. As expected,
in the instantaneously changing background case there is no evidence for a
new mode or nonquasinormal mode content, and the overtone is extracted cleanly. 

In Table~\ref{tab:ComparingBackgrounds}, we summarize the change in the amplitude
of the $(220)$ quasinormal measured at later times compared to the amplitude of the initial perturbation
for both initial data cases and for the fixed and evolving backgrounds.
As expected, we find a larger change when using initial data not corresponding
to a quasinormal mode of the final black hole. When using such initial data, 
there is a bigger change for the evolving background compared to the instantaneously changing
background. We conclude that the most conservative case to consider,
in terms of minimizing nonlinear effects, is that of a perturbation in the form of a remnant quasinormal mode on an evolving
background. 
\begin{center}
\begin{table}[ht]
\caption{Comparing initial data and background cases}
\label{tab:ComparingBackgrounds}
\begin{tabular}{ c|c|c } 
Initial data type & Background & $\Delta 220 / 220 $ \\
\hline
Remnant quasinormal mode & Evolving & 0.011 \\ 
Initial quasinormal mode & Evolving & 0.026 \\ 
Initial quasinormal mode & Instantaneous & 0.018 \\ 
\end{tabular}
\end{table}
\end{center}

Finally, we consider an instantaneously changing spherically symmetric background to compare to Ref.~\cite{Sberna:2021eui}.
There the amplitude of the inner product between the fundamental $200$ mode of a nonspinning black hole with mass $M$ and the overtone $201$ mode of a nonspinning black hole with mass $M+\Delta M$ was computed, 
giving $A^{(201)}_{\text{exc}}/A^{(200)}_{\text{pert}} = 6\times 10^{-3}$ for a mass change of $\Delta M/M = 10^{-3}$. 
We expect this calculation to give results equivalent to our mode analysis with background with mass $M+\Delta M$ and $a = 0$ and an initial perturbation corresponding to an $\ell = 2$, $m=0$, $n=0$ quasinormal mode
of a black hole with mass $M$.
We find, for a mass change of $\Delta M/M = 10^{-3}$, $A^{(201)}_{\text{exc}}/A^{(200)}_{\text{pert}} = 1.5 \times 10^{-3} $.
If we take the amplitudes from the start of the simulation in this case (without considering a lookback time), we find 
$
    A^{(201)}_{\text{exc}}/A^{(200)}_{\text{pert}} = 3 \times 10^{-3}.
$
This factor of a few difference could be due to some combination of coordinate differences and lookback time ambiguity.
\section{Numerical Convergence}
\label{sec:num_convergence}

\begin{figure}[h]
\centering
\includegraphics[width=0.5\textwidth]{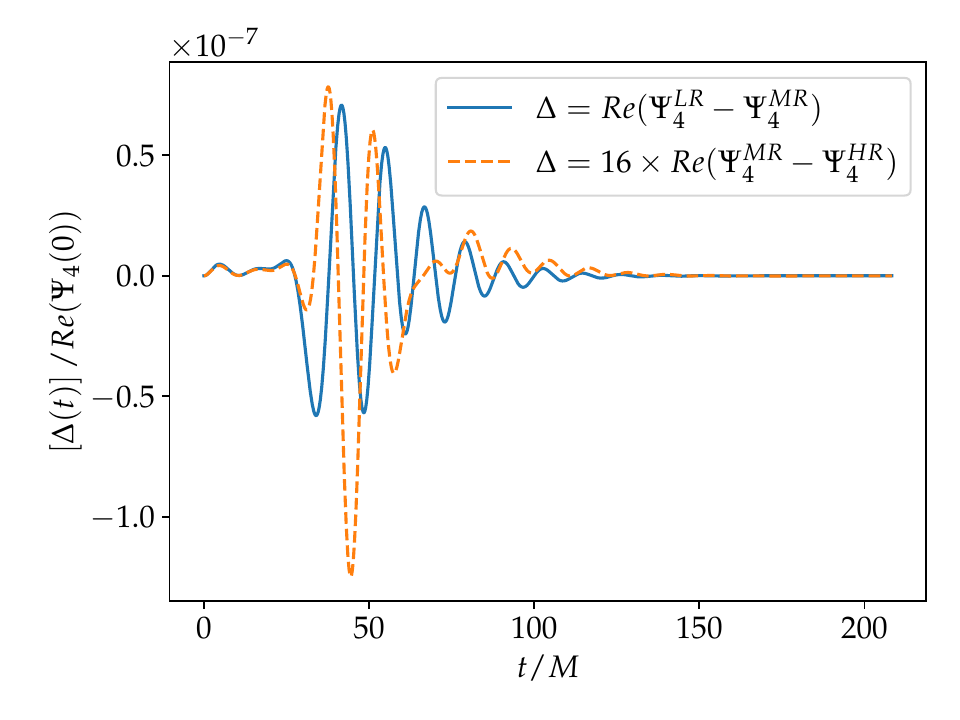}
\caption{Convergence plot showing the truncation error in the real part of $\Psi_4$ extracted at null infinity and projected onto the $\ell= m=2$ spin-weighted spherical harmonic. The factor of $16$ shown is consistent with fourth-order convergence. 
}
\label{fig:convergence_psi4}
\end{figure}

The evolution code uses a grid in real space in the radial direction, with
$n_R$ grid points, and a spectral grid in the angular directions, with $n_\ell$
spherical harmonics. The radial finite differencing is fourth order, as is the
time stepping. The standard choice for results included in the main text is
$n_R = 121$,\ $n_\ell = 25$.  We perform a pointwise convergence test on the
numerical evolution code with changing background in
Fig.~\ref{fig:convergence_psi4}. In this case, we consider three resolutions:
$n_R^{\text{LR}} = 121$, $n_R^{\text{MR}} = 241$, and $n_R^{\text{HR}} = 481$.
We scale the number of time steps with $n_R$, 
while keeping $n_\ell^{\text{LR}} = n_\ell^{\text{MR}} = n_\ell^{\text{HR}} =
25$ fixed. We find the expected fourth-order convergence, showing that the
radial/time resolution is dominating the total numerical error. We use this
scaling to calculate the numerical accuracy in the signal as a function of
time. We find that the relative error is $<10^{-5}$ for $t<100M$, 
though the relative error grows exponentially in time as the
quasinormal mode solutions decays exponentially.

We perform a convergence test for the extracted amplitudes $A_{\ell m
n}/\mathcal{A}$. These amplitudes are extracted by applying quasinormal mode
filters ~\eqref{eq:qnm_filter} to the data and then fitting in the time
domain for $t>10M$ with a fixed-frequency model
Eq.~\eqref{eq:fixed-freq-model}. We find that the fit amplitudes converge, with convergence order in the range 2-5 depending on the time of extraction. 
We check the convergence of extracted amplitudes for an evolving background in the cases of $a/M = 0.4$ and $a/M = 0.8$, as well as in the
instantaneously changing background and a remnant spin of $a/M=0.7$.
To calculate the numerical errors for fit amplitudes in Sec.~\ref{sec:Results}, we take two resolutions and assume only first-order convergence.

\bibliography{bib.bib}

\end{document}